%% file: main.tex
\documentclass{ceurart}%

\usepackage[acronym, nowarn]{glossaries}
\usepackage{url}
\makeglossaries
\loadglsentries{glossary}

\begin{document}

\copyrightyear{2021}
\copyrightclause{Copyright for this paper by its authors.
  Use permitted under Creative Commons License Attribution 4.0
  International (CC BY 4.0).}

\conference{3rd International Workshop and Challenge on Computer Vision in Endoscopy (EndoCV2021) \small{in conjunction with the 18th IEEE International Symposium on Biomedical Imaging ISBI2021,
  April 13th, 2021, Nice, France}}

\title{DivergentNets: Medical Image Segmentation by Network Ensemble}

\author[1,2]{Vajira Thambawita}[%
orcid=0000-0001-6026-0929,
email=Corresponding author: vajira@simula.no
]
\author[1,2]{Steven A. Hicks}[%
orcid=0000-0002-3332-1201
]
\author[1,2]{P{\aa}l Halvorsen}[%
orcid=0000-0003-2073-7029
]
\author[1]{Michael A. Riegler}[%
orcid=0000-0002-3153-2064
]

\address[1]{SimulaMet, Pilestredet 52, 0167 Oslo, Norway}
\address[2]{OsloMet, Pilestredet 46, 0167 Oslo, Norway}

\begin{abstract}
Detection of colon polyps has become a trending topic in the intersecting fields of machine learning and gastrointestinal endoscopy. The focus has mainly been on per-frame classification. More recently, polyp segmentation has gained attention in the medical community. Segmentation has the advantage of being more accurate than per-frame classification or object detection as it can show the affected area in greater detail. For our contribution to the EndoCV 2021 segmentation challenge, we propose two separate approaches. First, a segmentation model named \emph{TriUNet} composed of three separate UNet models. Second, we combine TriUNet with an ensemble of well-known segmentation models, namely UNet++, FPN, DeepLabv3, and DeepLabv3+, into a model called \emph{DivergentNets} to produce more generalizable medical image segmentation masks. In addition, we propose a modified Dice loss that calculates loss only for a single class when performing multi-class segmentation, forcing the model to focus on what is most important. Overall, the proposed methods achieved the best average scores for each respective round in the challenge, with {TriUNet being the winning model in Round I} and {DivergentNets being the winning model in Round II} of the {segmentation generalization challenge at EndoCV 2021}.
%
The implementation of our approach is made publicly available on GitHub.
\end{abstract}


\maketitle

\section{Introduction}
Automatic segmentation of medical images is a common use case in machine learning that has gained a lot of attention over the last few years. Popular applications include segmenting tumors in computed tomography (CT) scans~\cite{Alirr2020, ct_segment_pang}, finding abnormalities in magnetic resonance images (MRIs)~\cite{Yamanakkanavar2020, CLARKE1995343}, or segmenting organs and tissue in medical applications~\cite{Scheikl2020, Schoppe:2020aa}. Segmentation goes a step beyond standard classification and object detection as it extracts the area in an image that corresponds to the target class or classes at pixel-level precision. This comes with two advantages that are important in the medical field. The first one is that the algorithm learns pixel-wise and has more examples to learn from compared to if it would learn image-wise~\cite{Tiezhu, bansal2017pixelnet}. This can help for use cases where one does not have many images from a disease. Secondly, the segmented area makes it easier for the physician to determine what the algorithm detected and classified as a disease which serves in a broader sense as an explanation. Thus, detailed image segmentation can also be seen as a type of explanation method. This makes it highly desired by medical professionals as explainable machine learning is seen as one of the requirements for the successful implementation of automatic decision support systems in hospitals~\cite{amann2020explainability}.
\input{figures/image_examples}
As part of the EndoCV2021 challenge ({\url{https://endocv2021.grand-challenge.org/}}), we were tasked with creating machine learning models that automatically segment polyps~\cite{ali2021_endoCV2020,ali2020objective_EAD2019, jha2021real} in video frames collected from real-world endoscopies. This is a complex task as polyps come in various shapes and sizes, where some (e.g., flat lesions) are barely detectable by even the most experienced endoscopists. Figure~\ref{fig:image_examples} shows some of the more difficult examples taken from EndoCV's development dataset~\cite{polypGen2021} provided by the challenge organizers. The challenge presented two separate tasks, the \textit{detection generalization challenge} and the \textit{segmentation generalization challenge}. We participated in the segmentation generalization challenge, where we achieved the best results among 13 other competitors in both rounds. The code for the experiments presented in this paper is available on GitHub\footnote{\url{https://github.com/vlbthambawita/divergent-nets}}.

This paper summarizes our approaches to the EndoCV2021 challenge. In particular, we developed the \emph{TriUNet} segmentation model combining three separate UNet models, and the \emph{DivergentNets} that combines TriUNet with an ensemble of the well-known segmentation models, namely UNet++, FPN, DeepLabv3, and DeepLabv3+. The rest of the paper is structured as follows. Section~\ref{sec:approach} present our approach to this year's challenge, where we use two unique models that achieve state-of-the-art performance on the EndoCV dataset. Section~\ref{sec:experiments} gives a description of the implementation details on the models and training procedure and how the data was split and prepared. Section~\ref{sec:results} presents the preliminary and official results for our tested models and performs a qualitative analysis on some of the predicted masks. Lastly, Section~\ref{sec:conclusion} concludes this paper with a summary and plans for future work.
\section{Approach}\label{sec:approach}
In this section, we introduce three approaches that we developed for the segmentation generalization challenge at EndoCV 2021, which are two new architectures, TriUNet and DivergentNet, and a modified loss function.

\subsection{TriUNet}\label{sec:triunet}
TriUNet is a \gls{cnn} architecture that utilizes multiple UNet~\cite{Ronneberger2015} architectures arranged in a triangular structure as depicted in Figure~\ref{fig:TriUNet}. The model takes a single image as input, which is passed through two separate UNet models with different randomized weights. The output of both models is then concatenated before being passed through a third UNet model to predict the final segmentation mask. Figure~\ref{fig:TriUNet} also shows an example of the intermediate representations provided by the two initial UNet models. The loss is calculated and back-propagated through the whole model, meaning the entire network is trained in one go. From the intermediate representations, we clearly see that the different UNets learn different interpretations of the data, which then are combined in one final output.

\input{figures/triunet.tex}

\subsection{DivergentNets}
The DivergentNets network is inspired by the idea of ensembles made with multiple high-performing image segmentation architectures and the TriUNet architecture presented in the previous section. We constructed this DivergentNets assuming that cumulative decisions taken from multiple intermediate models should give a more precise decision than the predictions from a single network. The included models were selected based on what has previously been shown to produce good results on different segmentation tasks and some preliminary experiments using each model independently. Furthermore, the selection was limited by the hardware we had available.

As shown in Figure~\ref{fig:delphi_ensemble}, our configuration comprises five intermediate models, namely UNet++, FPN, DeepLabv3, DeepLabv3+, and TriUNet. The five intermediate models are first trained for $N$ number of epochs separately, where the best checkpoint of each model is selected to be combined in DivergentNet. This $N$ should be selected using a preliminary experiment. In our case, we identified that $N=200$ is enough to produce high-quality masks. However, training for more epochs may result in better checkpoints to use in DivergentNet. To produce the intermediate masks, the output of each model is passed through a \textit{softmax2d} activation function. However, this should be changed based on the application. In our case, we predict masks for two classes, background and polyp, where no two categories may overlap. The masks produced by each intermediate model represent the divergent views on the data. The final output of DivergentNets is made by averaging the pixels between each intermediate mask and rounding to the nearest integer (either 0 or 1).

\input{figures/delphi_ensemble}

\subsection{Single-channel Dice}\label{sec:dice_loss}
All models were trained to predict masks for both polyps and background (mostly containing the mucosal wall lining the inside of the colon). As the primary focus of EndoCV is to segment colon polyps, we use a modified Dice loss to calculate the prediction error. We call this loss function \textit{single-channel Dice loss} as it only considers one channel when calculating error. This is shown in Equation~\ref{eq:singledice}: 
\begin{equation}\label{eq:singledice}
    \text{Single-Channel Dice Loss} = \frac{2\cdot|A_n\cap B_n|}{2\cdot|A_n\cap B_n| + |B_n\backslash A_n| + |A_n\backslash B_n|}
\end{equation}
where $n$ represents the class for which loss should be calculated for. In this case, we only calculate loss for the polyp class and ignore the background.

\section{Experiments}\label{sec:experiments}
The experimenters can be categorized into two sub-groups, namely baseline experiments and experiments used for the challenge. The baseline experiments were used to benchmark common segmentation models. The baseline models tested were UNet~\cite{Ronneberger2015}, UNet++~\cite{zhou2018unet}, FPN~\cite{lin2017feature}, DeepLabv3~\cite{chen2017rethinking}, and DeepLabv3+~\cite{chen2018encoder}. In turn, we used these networks to design the TriUNet and DivergentNets architectures. This section describes the experimental setup, including how the data was prepared, training procedures, architecture implementations, and specifics on what hyperparameters were used.

\subsection{Data details and preparation}
The development dataset provided by the organizers was split between several directories, primarily one part consisting of a five-way center-wise split (directories \textit{C1} through \textit{C5}) containing a diverse set of data~\cite{polypGen2021}, and one part consisting of pure sequence data (directories \textit{seq1} through \textit{seq15}). For this challenge, we decided to use a standard three-way split of the data into training, validation, and testing datasets. The training data was made up of all the data contained within the center-wise split for training data, in addition to a few sequences only containing negative samples. For validation, we used the remaining sequence data. Table~\ref{tab:data} gives an overview of how each directory was split between training, validation, and test datasets. All samples contained an image, a segmentation mask, bounding box coordinates, and the image with the bounding-box superimposed over it. As we were only participating in the segmentation generalization challenge, we only used the images and segmentation masks.

\input{tables/datasets}

\subsection{Implementation details}
All models were implemented in PyTorch and trained on an Nvidia DGX-2. The Nvidia DGX-2 consists of 16 Tesla V100 GPUs, dual Intel Xeon Platinum 816 processors, and 1.5 terabytes of system memory. Despite that the system contains 16 GPUs, we only use one for training so that we can train multiple models in parallel. For the baseline experiments, we used the implementations and pre-trained weights available in the \textit{Segmentation Models}~\cite{Yakubovskiy:2019} library. These networks were also used as the basis for our proposed TriUNet and DivergentNets. Each model was implemented SE-ResNeXt-50-32x4D~\cite{hu2018squeeze} as the encoder, which was initialized with ImageNet~\cite{ImageNet} weights. Images and masks were resized to $256\times 256$ and resized back to the original resolution using bilinear interpolation. The final prediction was produced by passing the output through a two-dimensional softmax function. For training, all models started with a learning rate of $0.0001$ and reduced to $0.00001$ after $50$ epochs. The model error was calculated using the proposed single-channel Dice for the polyp class (as explained in Section~\ref{sec:dice_loss}), and the weights were optimized using Adam~\cite{kingma2014adam}.

As the size of the development dataset is relatively small, we use a series of different image augmentations to make the model more generalizable. These augmentations include horizontal flip, shift scale rotation, resizing, additive Gaussian noise, perspective shift, contrast limited adaptive histogram equalization (CLAHE), random brightness, random gamma, random sharpen, random blur, random motion blur, random contrast, and hue saturation. The augmentations were implemented using the Python library \textit{Albumentations}~\cite{buslaev2020albumentations}. No augmentations were applied to the validation and testing data.

\section{Results and Discussion}\label{sec:results}
In this section, we discuss the preliminary and official results of our approach to the EndoCV 2021 challenge. We also perform a qualitative analysis of the models, showing how the different modes diverge to a final prediction.

\subsection{Preliminary results}

\input{tables/validation_results}
\input{tables/test_results}

Table~\ref{tab:valid_results} and Table~\ref{tab:test_results} show the initial results on the provided development validation and testing datasets. Overall, we see that all models perform well on segmenting the polyp class, with the DivergentNets architecture achieving the best performance and UNet++ at a close second place on both the validation and test datasets. Comparing UNet and TriUNet, we see that TriUNet performs slightly better on the polyp class, however, UNet++ outperforms both. With these results, it would be natural to assume that a TriUNet++ architecture would perform even better than TriUNet. However, due to hardware limitations (specifically GPU memory), we were unable to test this configuration and move this to future work.
\subsection{Official results}
\input{tables/official_results}
The official evaluation was split into two rounds, where \textit{Round I} used a subset of the testing data that was fully used for \textit{Round II}. For both rounds, we were limited by the number of submissions that could be delivered per day. This limit started at five-per-day for \textit{Round I} and was reduced to two-per-day for \textit{Round II}. Due to this limitation, only a subset of the aforementioned models was submitted as official runs. Models were selected based on their performance on a different test dataset that we chose, namely the well-known and established HyperKvasir~\cite{borgli2020hyperkvasir} dataset, in ascending fashion. From the HyperKvasir dataset, we only used the images with segmentation masks as an independent test set to determine the best generalizable model. It was not used in any way as training or validation data. Table~\ref{tab:official_results} shows the official results for \textit{Round I} and \textit{Round II}. Note that the DivergentNets model was not part of \textit{Round I} as it was developed during \textit{Round I} and used in \textit{Round II} once it was finished. From the results, we see that TriUNet achieved the best score for \textit{Round I}, and DivergentNets achieved the best score for \textit{Round II}, \emph{i.e.},{both winning their respective rounds of the competition}.

\subsection{Qualitative analysis}
\input{figures/predicted_masks}
Figure~\ref{fig:predicted_masks} shows some example masks predicted by our best performing model (DivergentNets) together with masks produced by the intermediate models. We see that each intermediate model learns slightly different features, making an overall more precise segmentation mask when combined. For example, the first row of Figure~\ref{fig:predicted_masks} shows the predicted masks and ground truth of a large polyp. We see that each model predicts slightly different masks for the same input and that TriUNet over-estimates the size of the polyp. After averaging the predicted masks for DivergentNets' final output, this area is smoothed out by the predictions from the other intermediate models. 

Even though DivergentNets primarily produces more accurate masks than any single model, there are cases where masks from the intermediate model better match the ground truth. We see this in row three, where DeepLabv3+ produces a more precise mask than all other intermediate models, making the averaged output less accurate.
\section{Conclusion and future work}\label{sec:conclusion}
In this paper, we presented our approaches to the EndoCV 2021 challenge. We trained a series of baseline models and two models based on novel architectures using a slightly modified Dice loss, which achieved the overall best score in both rounds of the generalization segmentation challenge. For the first round, we developed TriUNet, which reached an average score of $0.925$ on the official testing dataset. For the second round, we developed the DivergentNets architecture, which combines the baseline models with the TriUNet to gain an average score of $0.823$ on the official training dataset. Due to a limitation on time and computational resources, we could not experiment with another improved version where the UNet architectures are replaced with UNet++ architectures. 

For future work, we plan to explore different configurations of TriUNet, such as implementing TriUNet++ and testing different architectures for three architectures that make up the TriUNet architecture, for example, combining the UNet, FPN, and DeepLabv3 as TriUNet nodes. We would also like to explore different configurations for the DivergentNets architecture with different networks for each node. Another idea could be to use a neural network to produce the final prediction instead of the current averaging technique, similar to the approach discussed in~\cite{10.1145/3386295}. Further testing the approaches with datasets from other medical fields can help to identify the generalizability of our approach.
\section*{Acknowledgments}
The research has benefited from the Experimental Infrastructure for Exploration of Exascale Computing (eX3), which is financially supported by the Research Council of Norway under contract 270053.
\bibliography{ref}
\end{document}

%% file: glossary.tex
\newacronym{cnn}{CNN}{convolutional neural network}
\newacronym{ct}{CT}{computed tomography}
\newacronym{mri}{MRI}{magnetic resonance image}

%% file: figures/image_examples.tex
\begin{figure}[t!]
    \newcommand{\imagesize}{.16\linewidth}
    \centering
    \includegraphics[width=\imagesize, height=\imagesize]{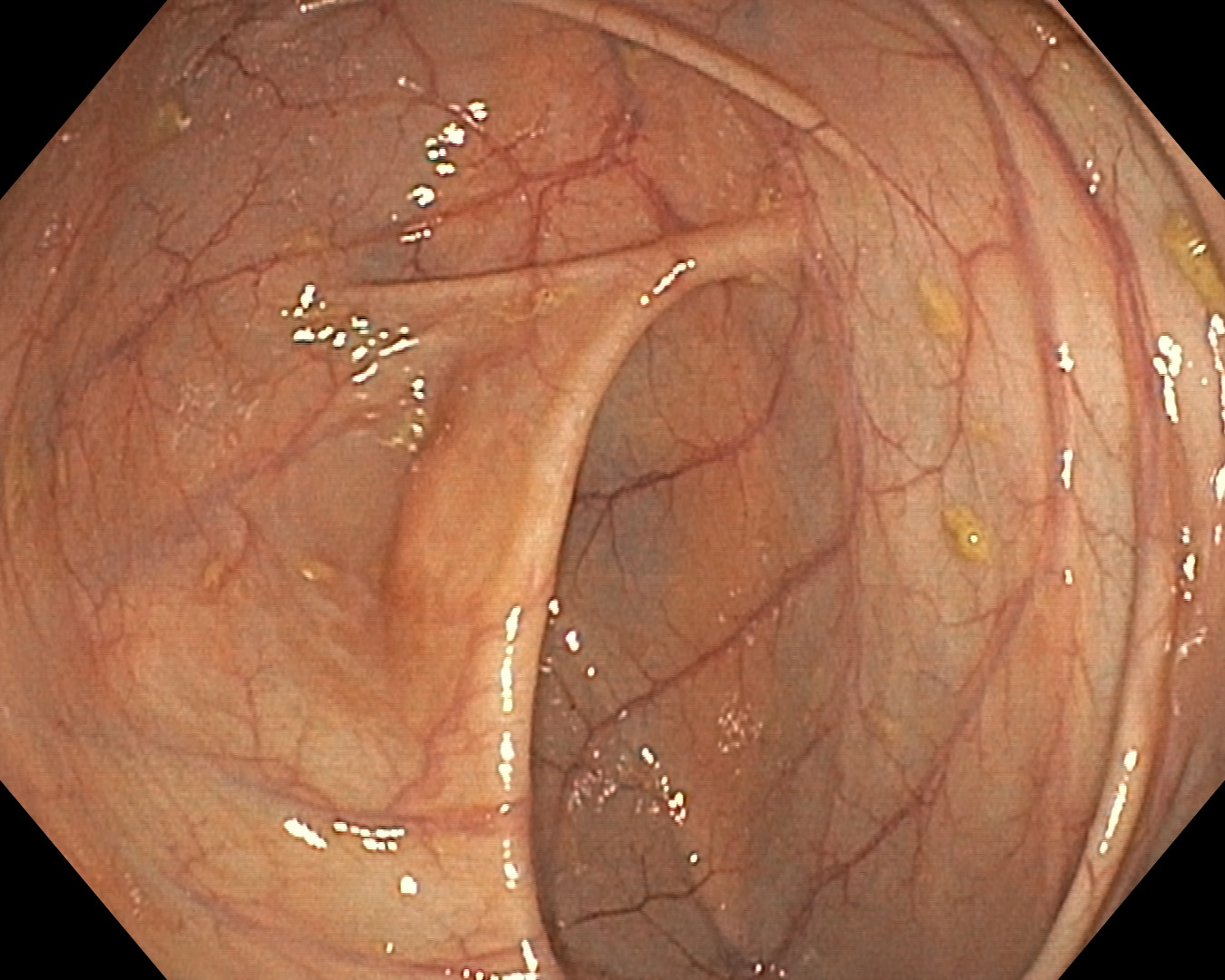}
    \includegraphics[width=\imagesize, height=\imagesize]{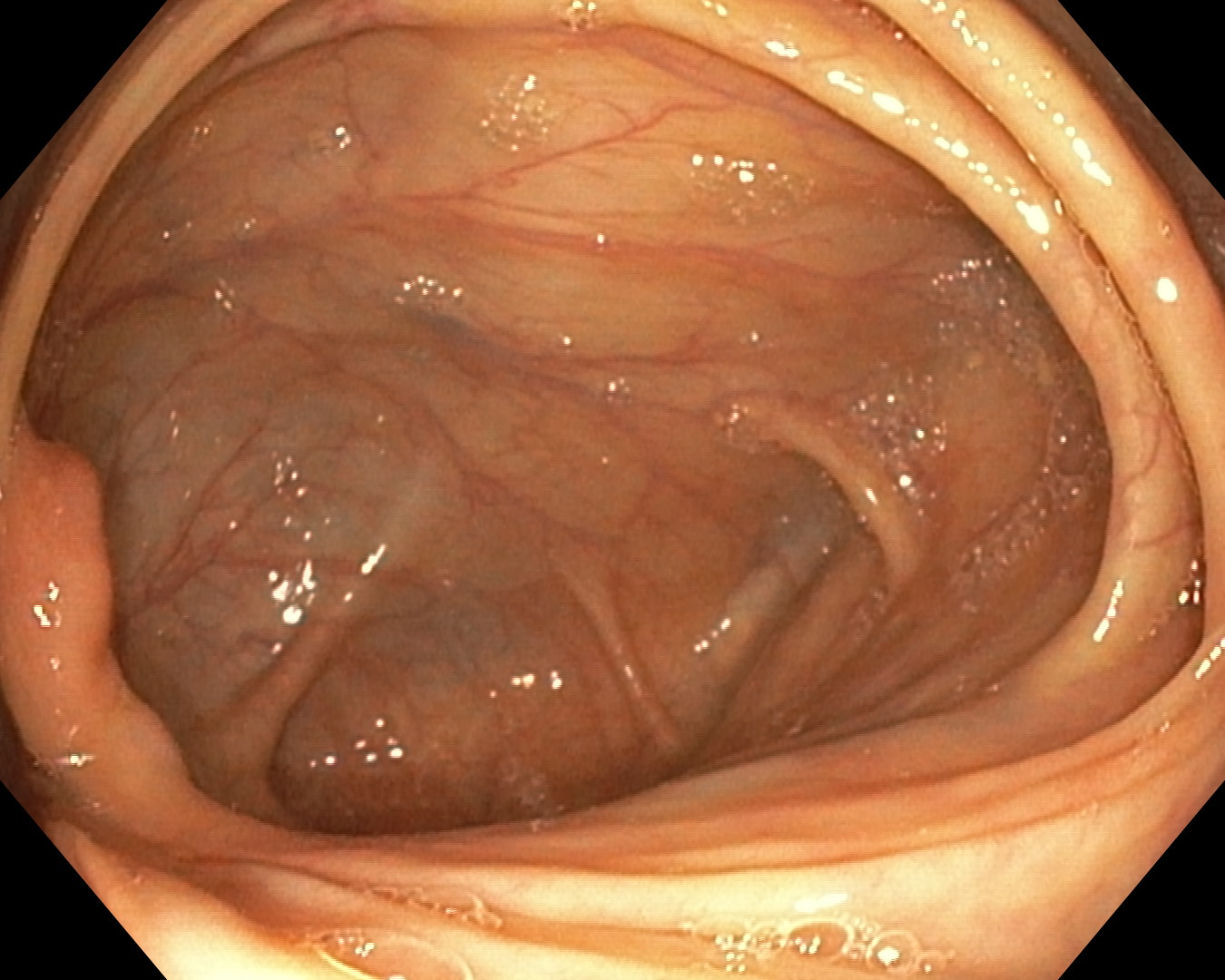}
    \includegraphics[width=\imagesize, height=\imagesize]{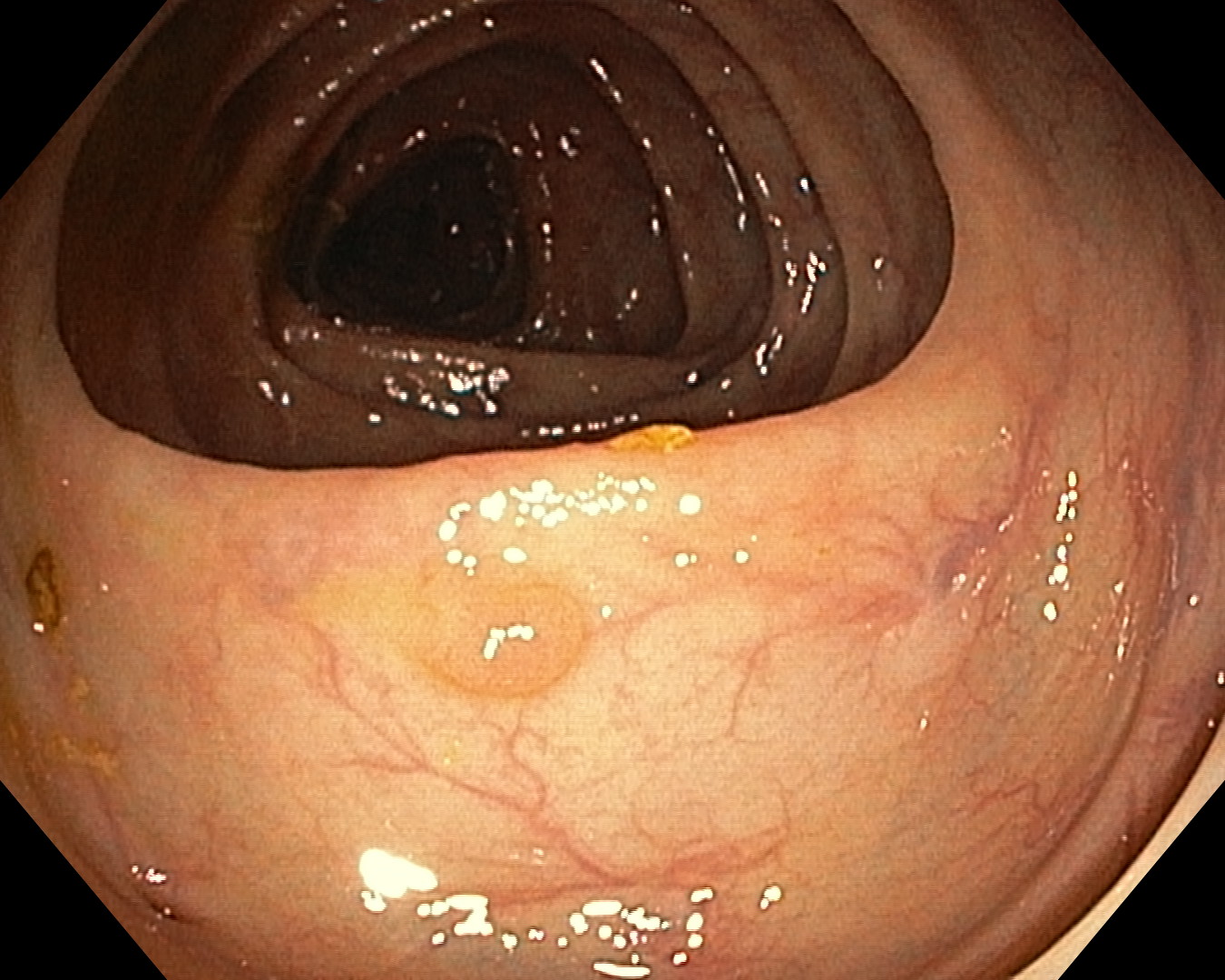}
    \includegraphics[width=\imagesize, height=\imagesize]{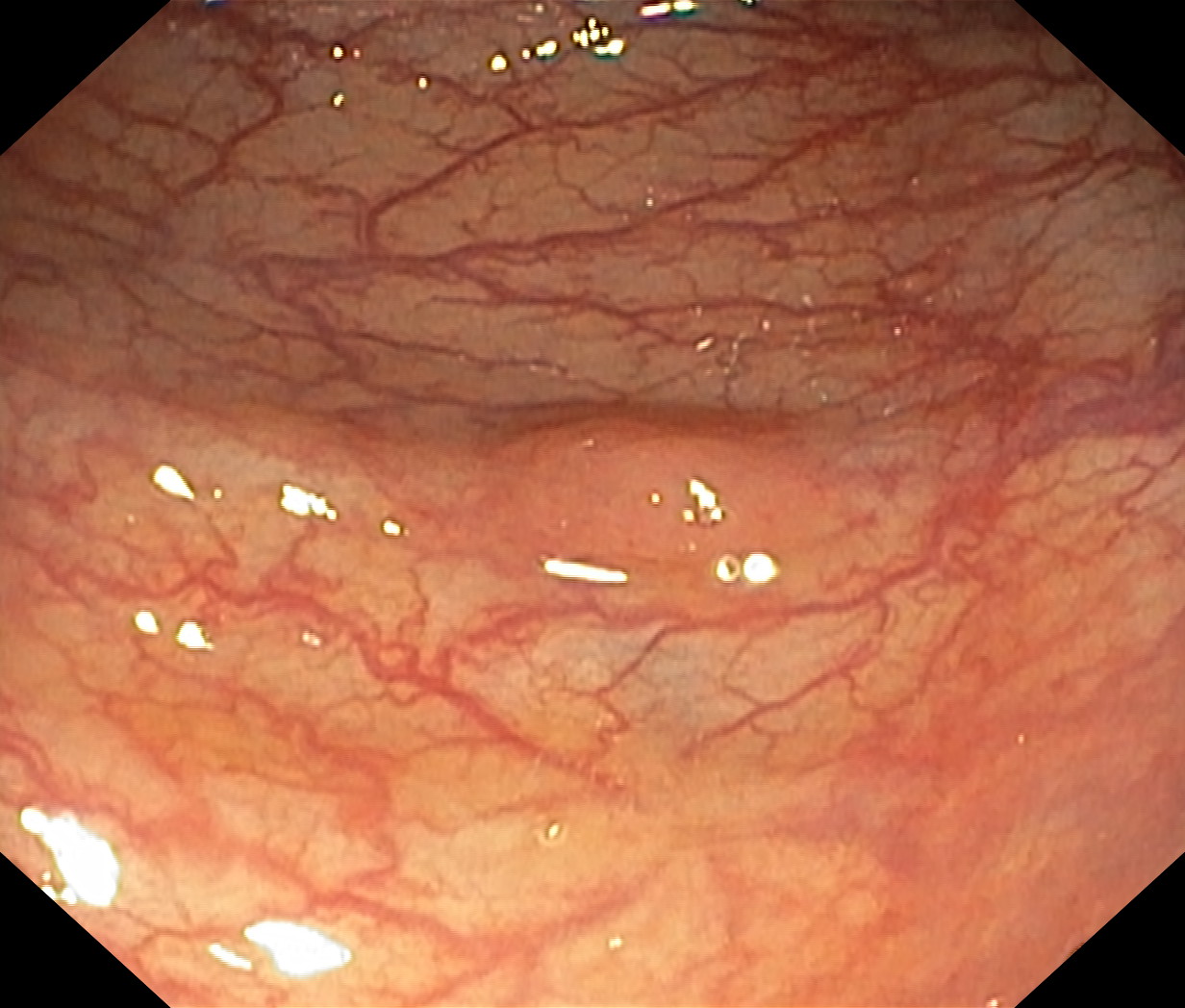}
    \includegraphics[width=\imagesize, height=\imagesize]{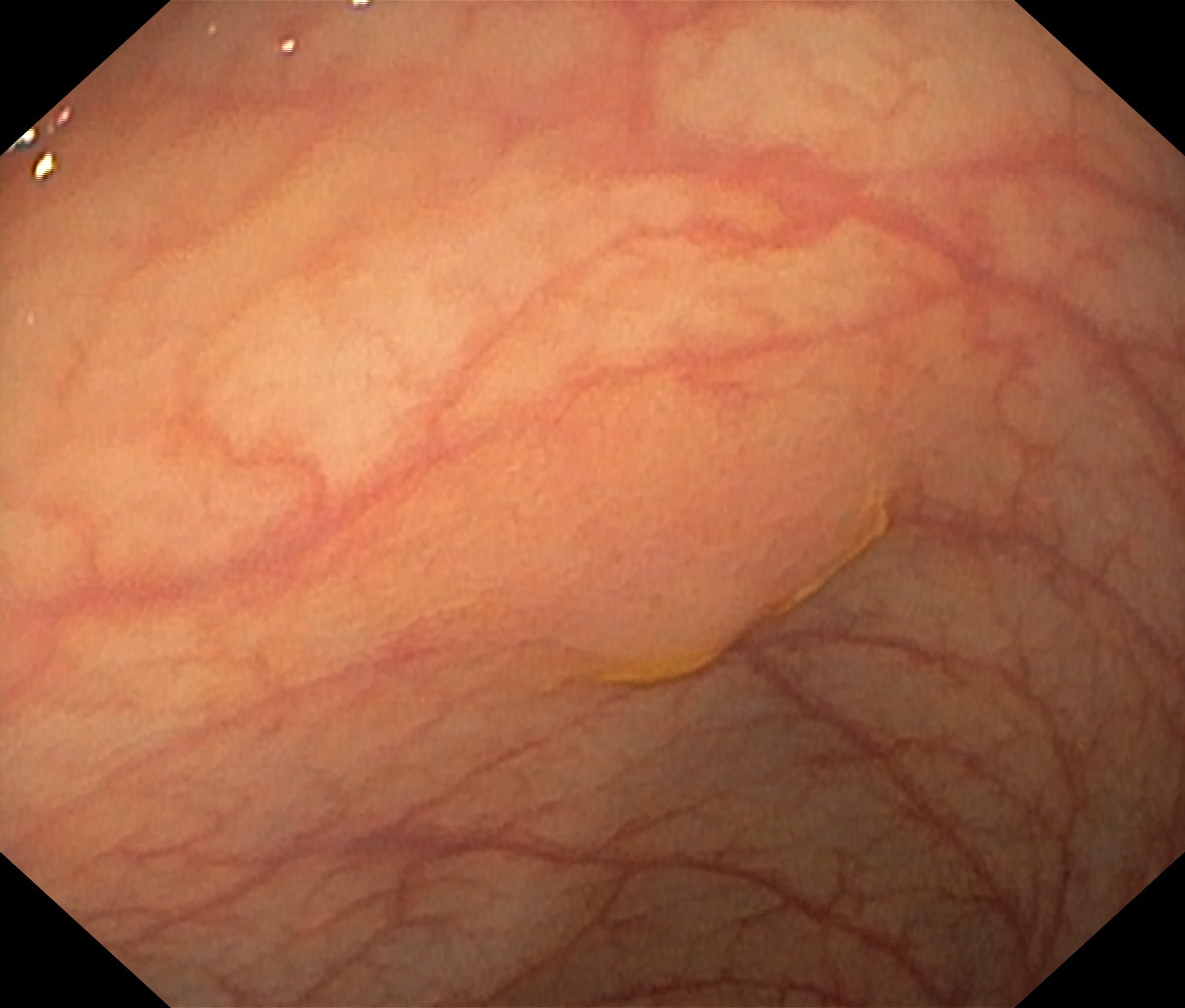}
    \includegraphics[width=\imagesize, height=\imagesize]{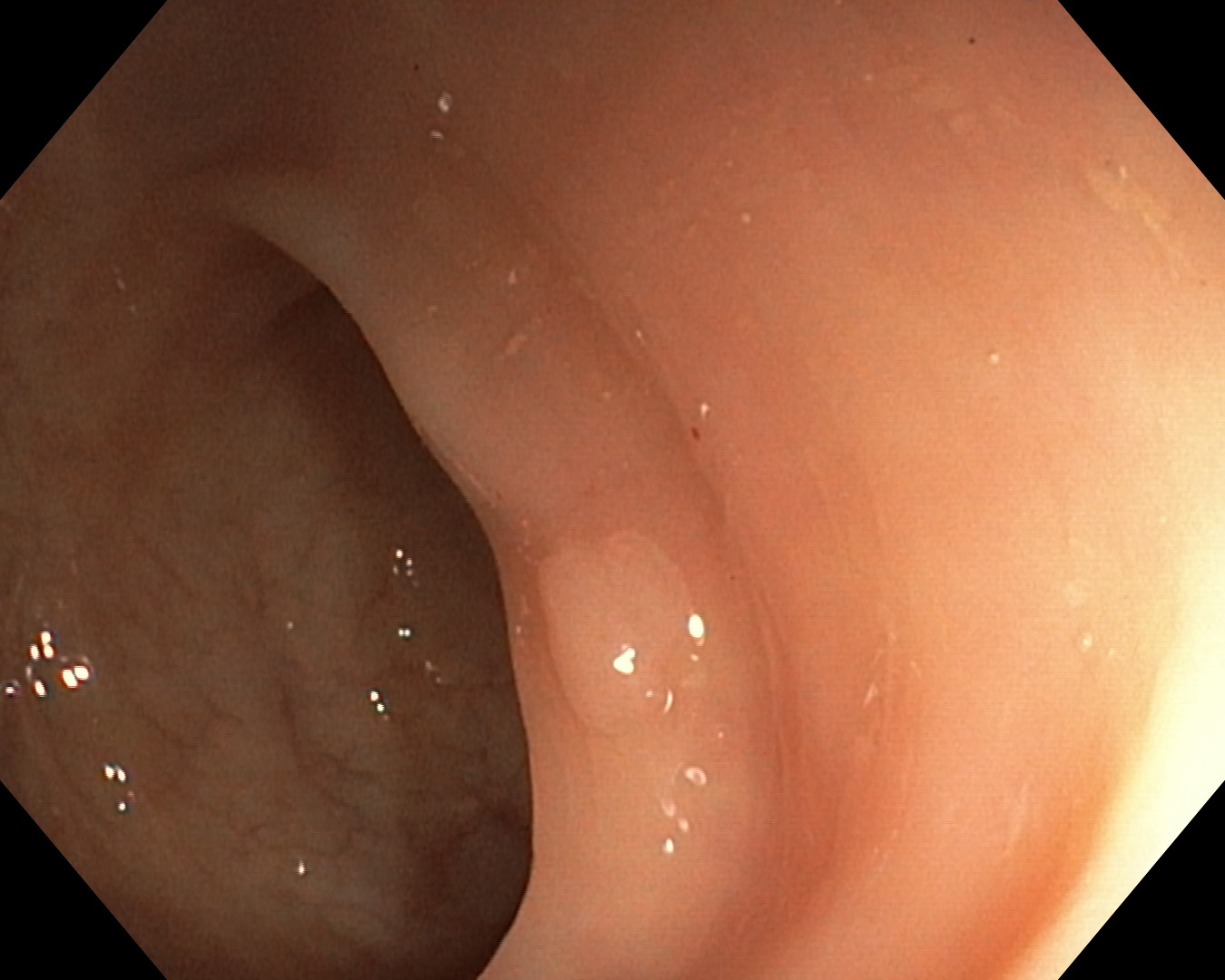}
    
    \includegraphics[width=\imagesize, height=\imagesize]{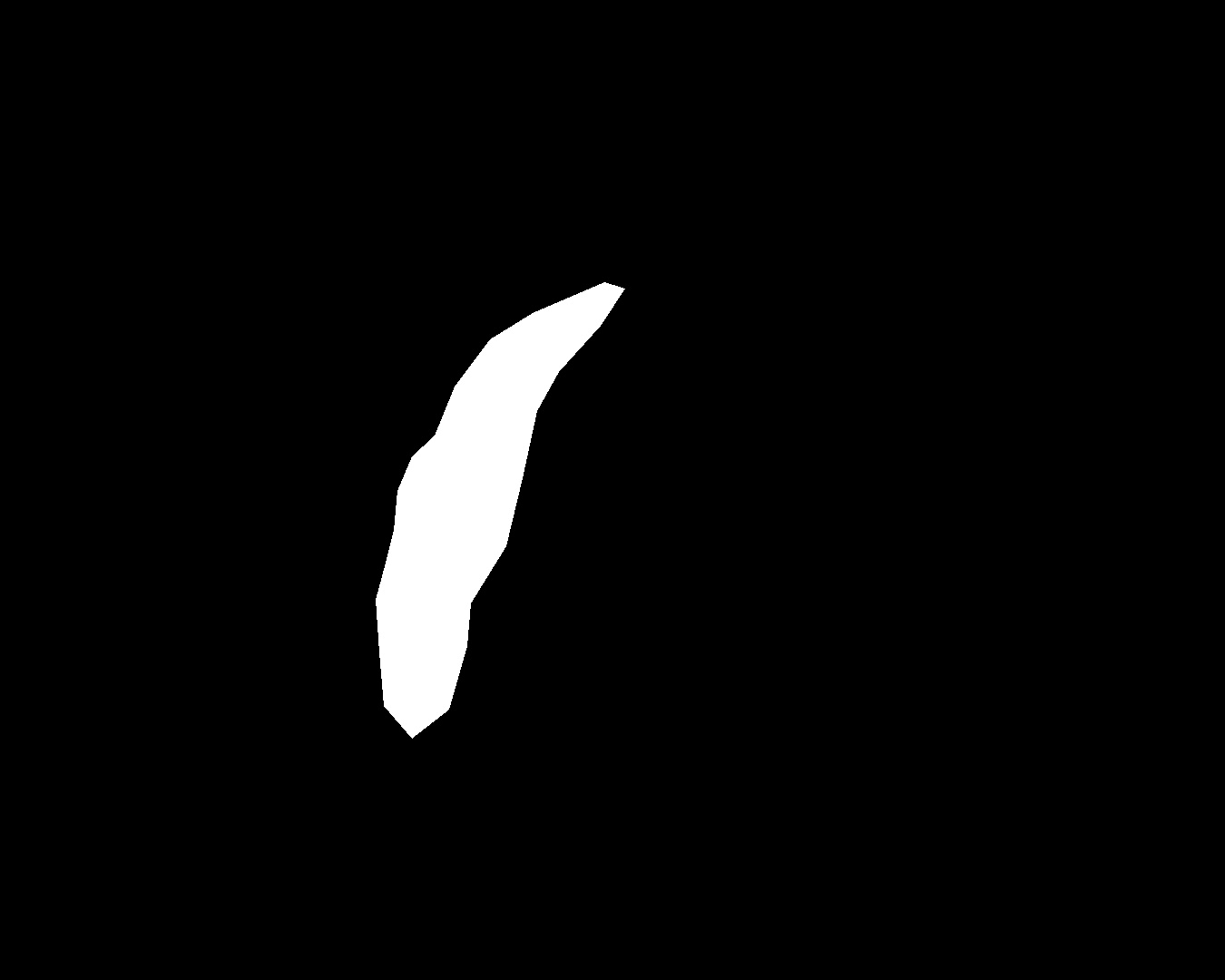}
    \includegraphics[width=\imagesize, height=\imagesize]{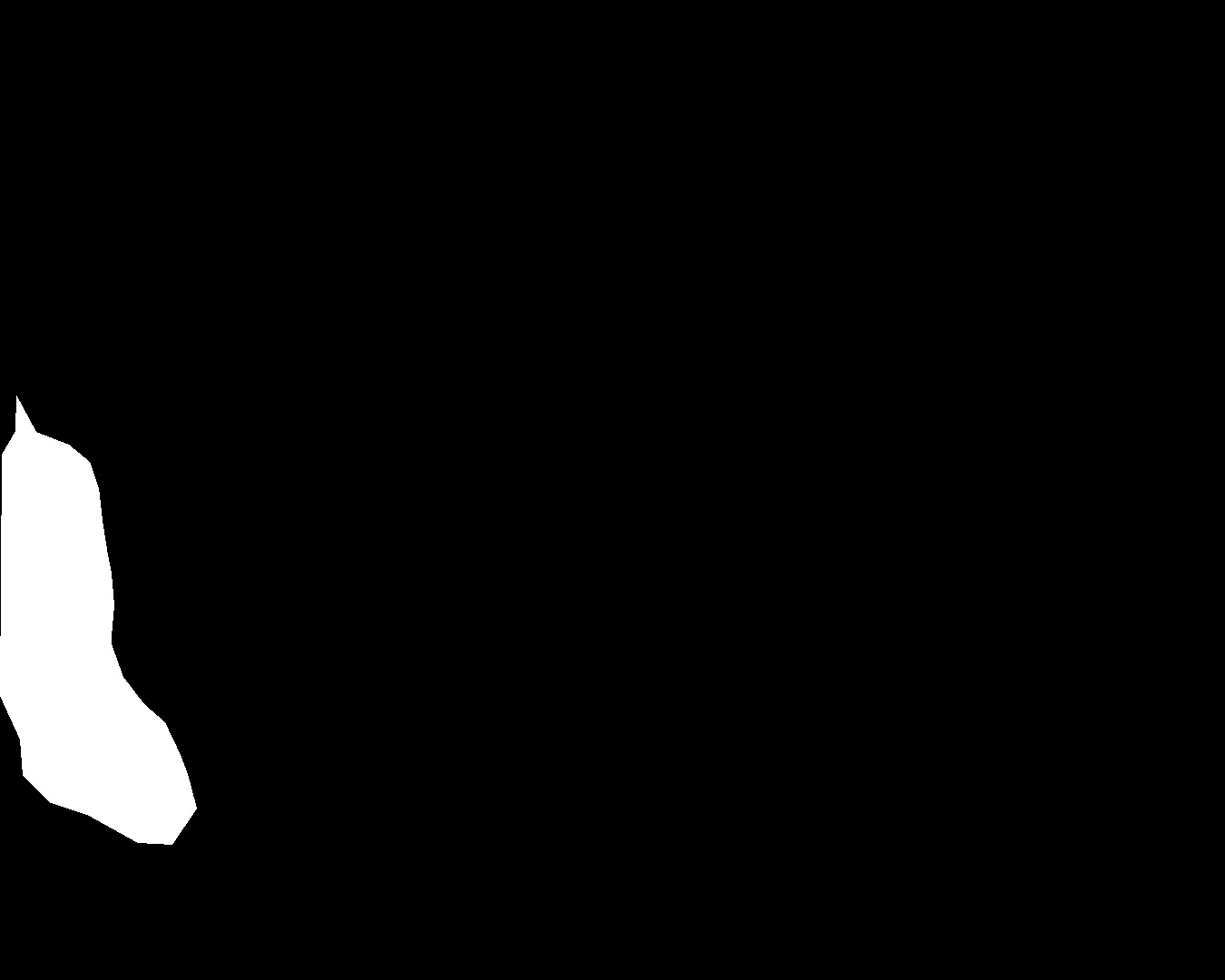}
    \includegraphics[width=\imagesize, height=\imagesize]{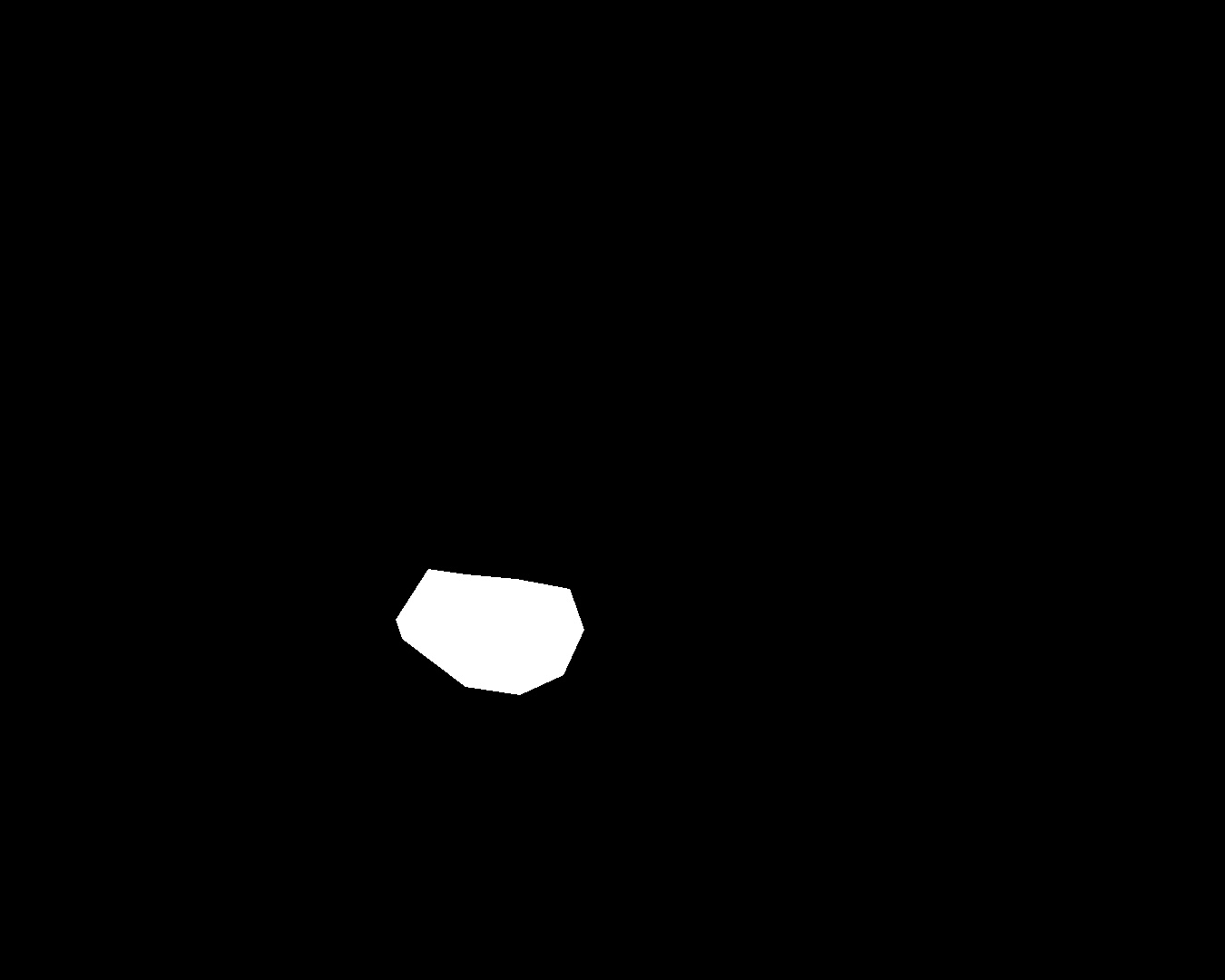}
    \includegraphics[width=\imagesize, height=\imagesize]{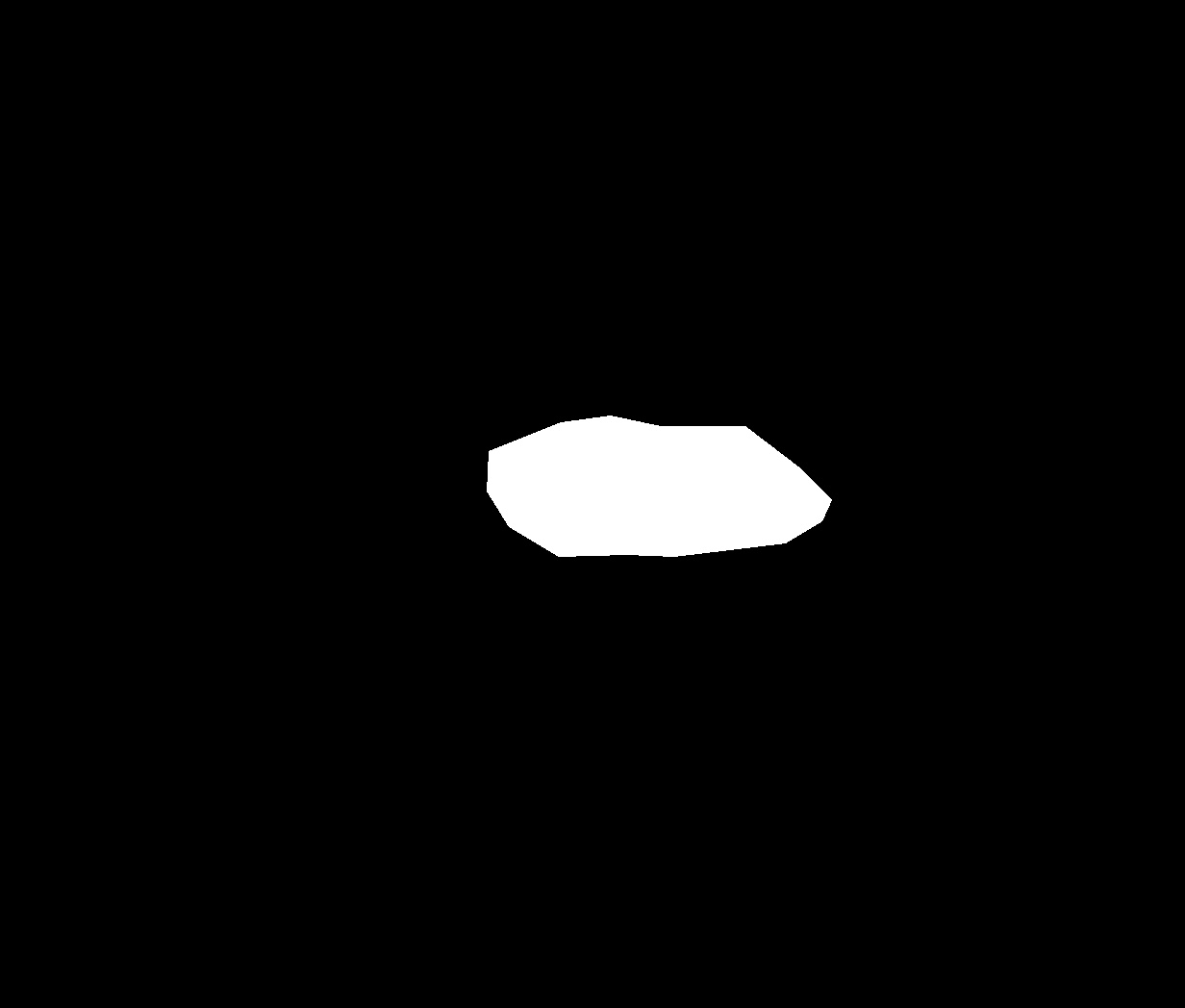}
    \includegraphics[width=\imagesize, height=\imagesize]{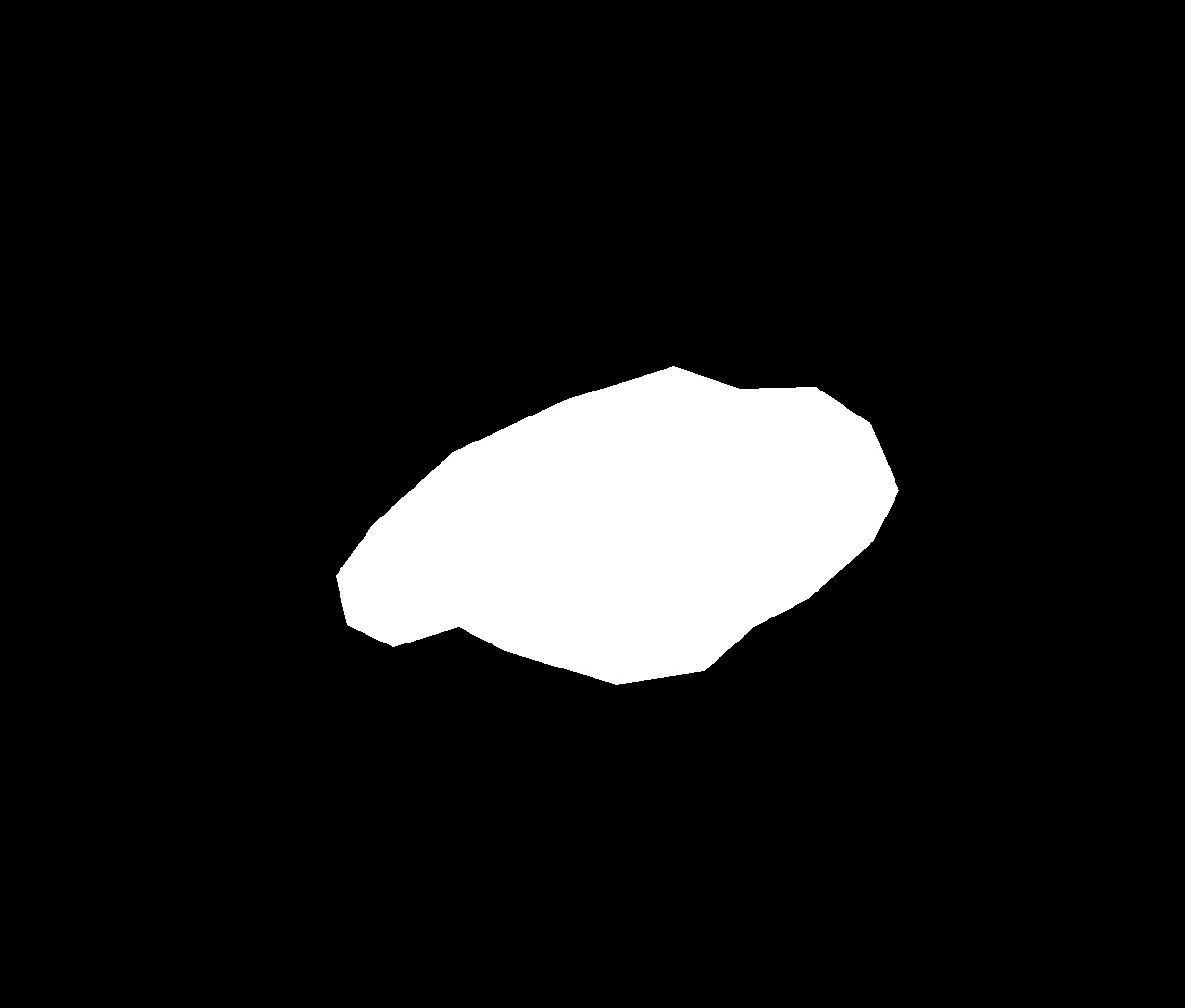}
    \includegraphics[width=\imagesize, height=\imagesize]{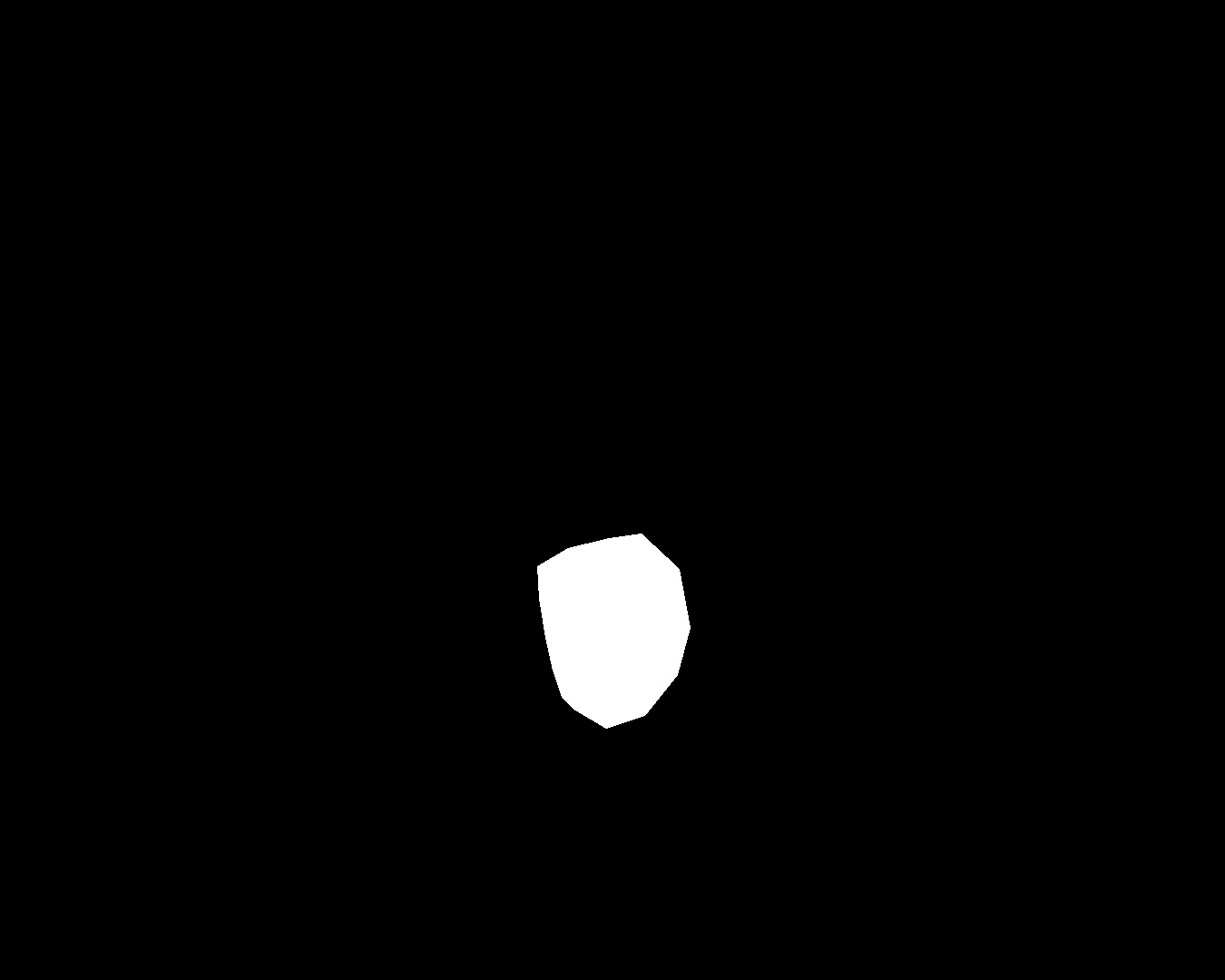}
    
    \caption{Some example images and their corresponding ground truth masks taken from the development dataset of EndCV2021~\cite{polypGen2021}}
    \label{fig:image_examples}
\end{figure}

%% file: figures/triunet.tex
\begin{figure}[t!]
    \centering
    \includegraphics[width=\textwidth]{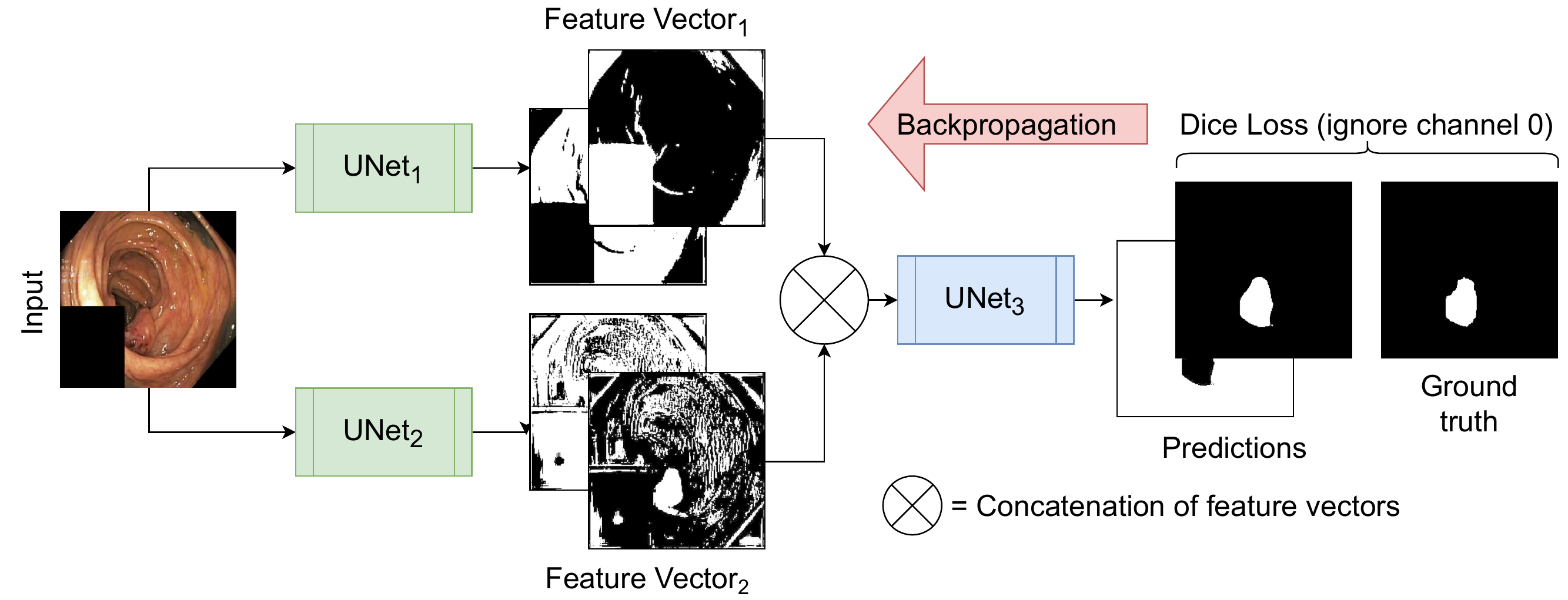}
    \caption{An illustration of the TriUNet architecture. First, the image is passed through two separate UNets in parallel, which produce the feature vectors $V_1$ and $V_2$, respectively. These two vectors are then concatenated before being passed through a third UNet that predicts the final segmentation mask. The loss is calculated by taking the Dice coefficient of the mask corresponding to the main class and the ground truth, which is then back-propagated through the entire model.}
    \label{fig:TriUNet}
\end{figure}

%% file: figures/delphi_ensemble.tex
\begin{figure}[t!]
    \centering
    \includegraphics[width=0.95\textwidth]{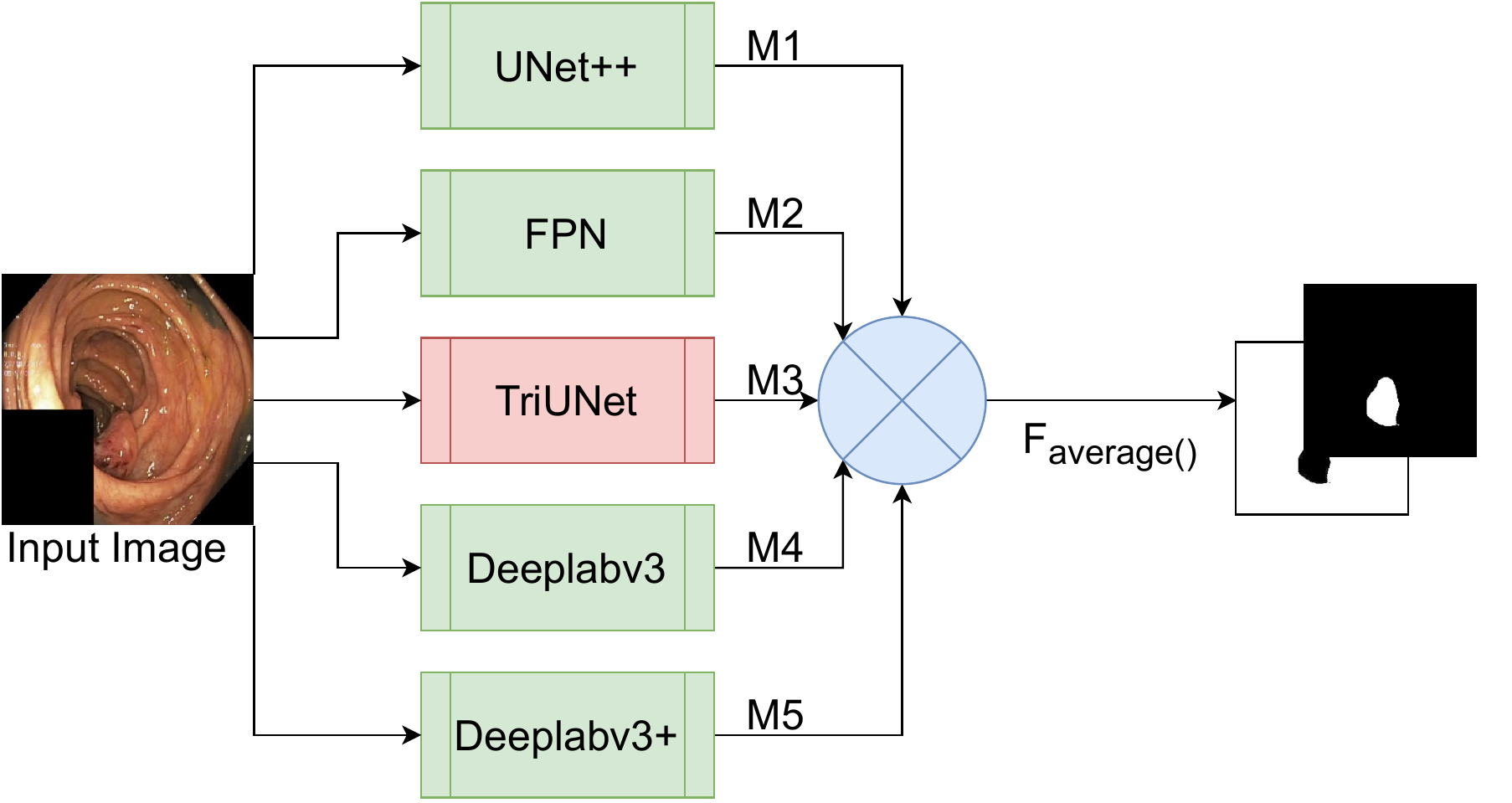}
    \caption{An illustration of the DivergentNets architecture. First, five different models are trained using the U-Net++, FPN, TriUnet, DeepLabv3, and DeepLabv3 architectures. Then, an image is passed through each model separately, which produces masks $M_1$ to $M_5$. Last, the masks are averaged to make the final segmentation mask.}
    \label{fig:delphi_ensemble}
\end{figure}

%% file: tables/datasets.tex
\begin{table}
    \centering
    \caption{An overview of how the data was split between training, validation, and testing.}
    \scriptsize
    \begin{tabular}{ r r r r r r }
        \toprule
        Dataset & Partition & \# Samples & \# Polyp & \# Non-Polyp \\
        \midrule
        EndoCV & Training & 1,754 & 1,329 & 435 \\
        EndoCV & Validation & 2,756 & 1,400 & 1,356 \\
        HyperKvasir & Testing & 1,000 & 1,000 & 0 \\
        \bottomrule
    \end{tabular}
    \label{tab:data}
\end{table}

%% file: tables/validation_results.tex
\begin{table}
    \centering
    \scriptsize
    \caption{The results collected from the preliminary experiments on the internal validation dataset.}\label{tab:valid_results}
    \begin{tabular}{ r r c c c c c c c c c c c c }
        \toprule
        \multirow{2}{*}{Model} & \multicolumn{4}{c}{All Classes} & \multicolumn{4}{c}{Polyp Class} & \multicolumn{4}{c}{Background Class} \\
        & IoU & F1 & REC & PREC & IoU & F1 & REC & PREC & IoU & F1 & REC & PREC \\
        \cmidrule(lr){2-5}\cmidrule(lr){6-9}\cmidrule(lr){10-13}
        U-Net      & 0.973 & 0.985 & 0.985 & 0.985 & 0.774 & 0.802 & 0.831 & 0.926 & 0.984 & 0.991 & 0.996 & 0.988 \\
        U-Net++    & 0.972 & 0.984 & 0.984 & 0.984 & 0.787 & 0.815 & 0.847 & 0.918 & 0.983 & 0.991 & 0.995 & 0.989 \\
        FPN        & 0.973 & 0.985 & 0.985 & 0.985 & 0.778 & 0.810 & \textbf{0.853} & 0.904 & 0.984 & 0.991 & 0.995 & 0.989 \\
        DeepLabv3  & 0.971 & 0.984 & 0.984 & 0.984 & 0.764 & 0.798 & 0.842 & 0.902 & 0.983 & 0.991 & 0.994 & 0.989 \\
        DeepLabv3+ & 0.973 & 0.985 & 0.985 & 0.985 & 0.777 & 0.807 & 0.840 & 0.919 & 0.984 & 0.991 & 0.994 & 0.989 \\
        TriUNet   & 0.970 & 0.983 & 0.983 & 0.983 & 0.775 & 0.802 & 0.846 & 0.903 & 0.982 & 0.990 & 0.992 & \textbf{0.989} \\
        DivergentNets     & \textbf{0.976} & \textbf{0.986} & \textbf{0.986} & \textbf{0.986} & \textbf{0.795} & \textbf{0.823} & 0.844 & \textbf{0.937} & \textbf{0.986} & \textbf{0.992} & 0.997 & 0.989 \\
        \bottomrule
    \end{tabular}
\end{table}

%% file: tables/test_results.tex
\begin{table}
    \centering
    \scriptsize
    \caption{The results collected from the preliminary experiments on the internal testing dataset.}\label{tab:test_results}
    \begin{tabular}{ r r c c c c c c c c c c c c }
        \toprule
        \multirow{2}{*}{Model} & \multicolumn{4}{c}{All Classes} & \multicolumn{4}{c}{Polyp Class} & \multicolumn{4}{c}{Background Class} \\
        & IoU & F1 & REC & PREC & IoU & F1 & REC & PREC & IoU & F1 & REC & PREC \\
        \cmidrule(lr){2-5}\cmidrule(lr){6-9}\cmidrule(lr){10-13}
        U-Net      & 0.941 & 0.967 & 0.967 & 0.967 & 0.823 & 0.883 & 0.876 & 0.938 & 0.959 & 0.977 & 0.988 & 0.970 \\
        U-Net++    & 0.945 & 0.969 & 0.969 & 0.969 & 0.834 & 0.894 & 0.882 & 0.942 & 0.961 & 0.979 & 0.988 & 0.972 \\
        FPN        & 0.944 & 0.968 & 0.968 & 0.968 & 0.824 & 0.887 & 0.870 & 0.943 & 0.961 & 0.978 & \textbf{0.990} & 0.970 \\
        DeepLabv3  & 0.942 & 0.968 & 0.968 & 0.968 & 0.821 & 0.885 & 0.874 & 0.935 & 0.959 & 0.977 & 0.988 & 0.970 \\
        DeepLabv3+ & 0.942 & 0.968 & 0.968 & 0.968 & 0.823 & 0.886 & 0.883 & 0.931 & 0.823 & 0.886 & 0.883 & 0.931 \\
        TriUNet   & 0.941 & 0.967 & 0.967 & 0.967 & 0.829 & 0.890 & \textbf{0.891} & 0.928 & 0.959 & 0.977 & 0.983 & \textbf{0.975} \\
        DivergentNets     & \textbf{0.949} & \textbf{0.972} & \textbf{0.972} & \textbf{0.972} & \textbf{0.840} & \textbf{0.899} & 0.886 & \textbf{0.946} & \textbf{0.964} & \textbf{0.980} & 0.990 & 0.973 \\
        \bottomrule
    \end{tabular}
\end{table}

%% file: tables/official_results.tex
\begin{table}
    \centering
    \scriptsize
    \caption{The official results provided by the EndoCV organizers. \textit{Score} is an average score of F1-score, F2-score, PPV, and Recall provided by the organizers, and \textit{SD} is the standard deviation of the metrics.}\label{tab:official_results}
    \begin{tabular}{ c c c c }
        \toprule
        Round & Model & Score & SD \\
        \midrule
        \multirow{2}{*}{I}  & UNet++    & 0.917 & 0.168 \\
                            & TriUNet   & \textit{0.925} & \textit{0.152} \\
        \midrule
        \multirow{2}{*}{II} & TriUNet   & 0.796 & 0.047 \\
                            & DivergentNets & \textit{0.823} & \textit{0.043} \\
        \bottomrule
    \end{tabular}
\end{table}

%% file: figures/predicted_masks.tex
\begin{figure}[!t]
    \newcommand{\imagesize}{.12}
    \scriptsize
    \setlength{\tabcolsep}{1pt}

\begin{tabular}{ c c c c c c c c}
    
    Image & DeepLabv3 & DeepLabv3+ & FPN & U-Net++ & TriUNet & DivergentNets & Ground Truth \\
    
    \includegraphics[width=\imagesize\linewidth,height=\imagesize\linewidth]{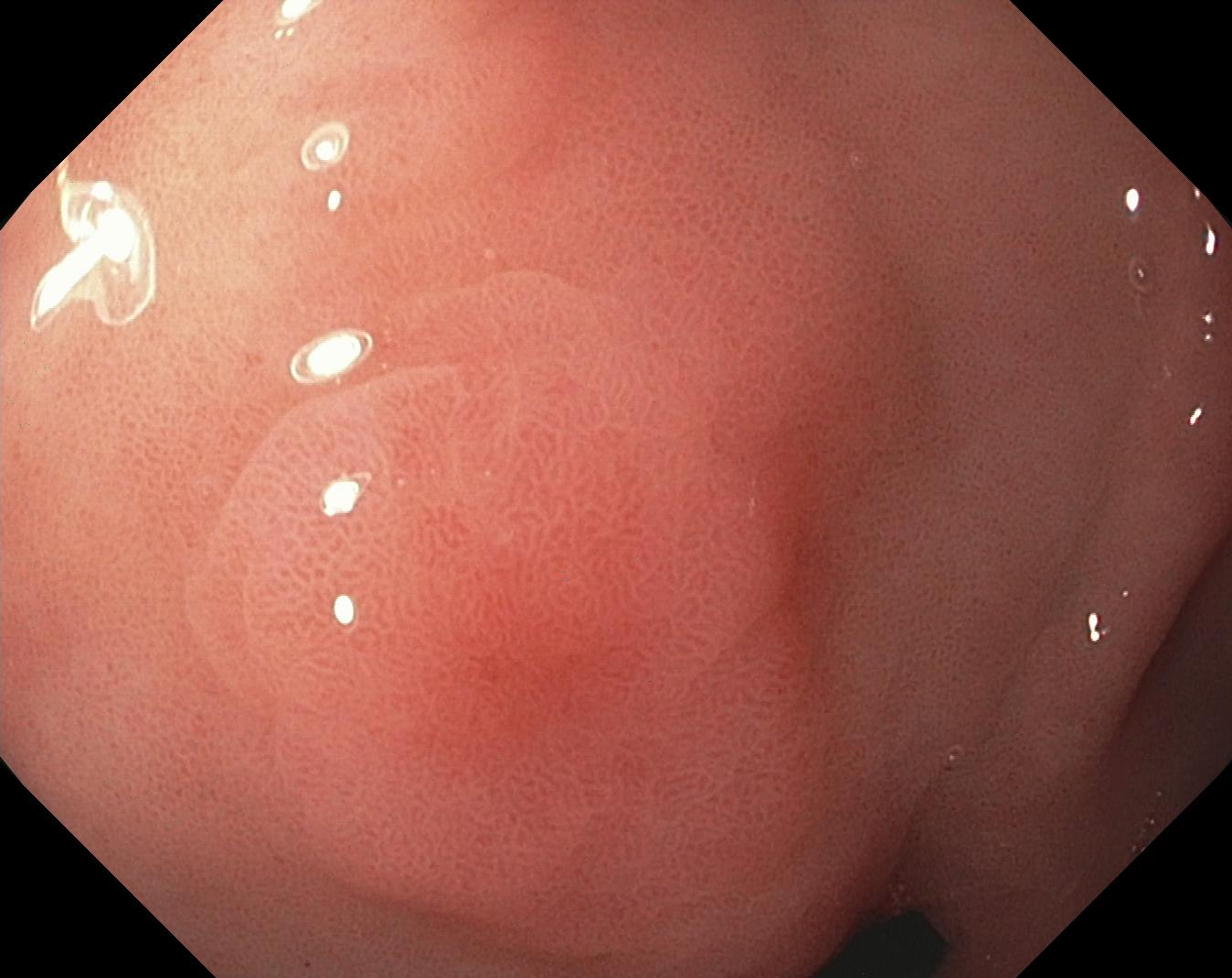} &
    \includegraphics[width=\imagesize\linewidth,height=\imagesize\linewidth]{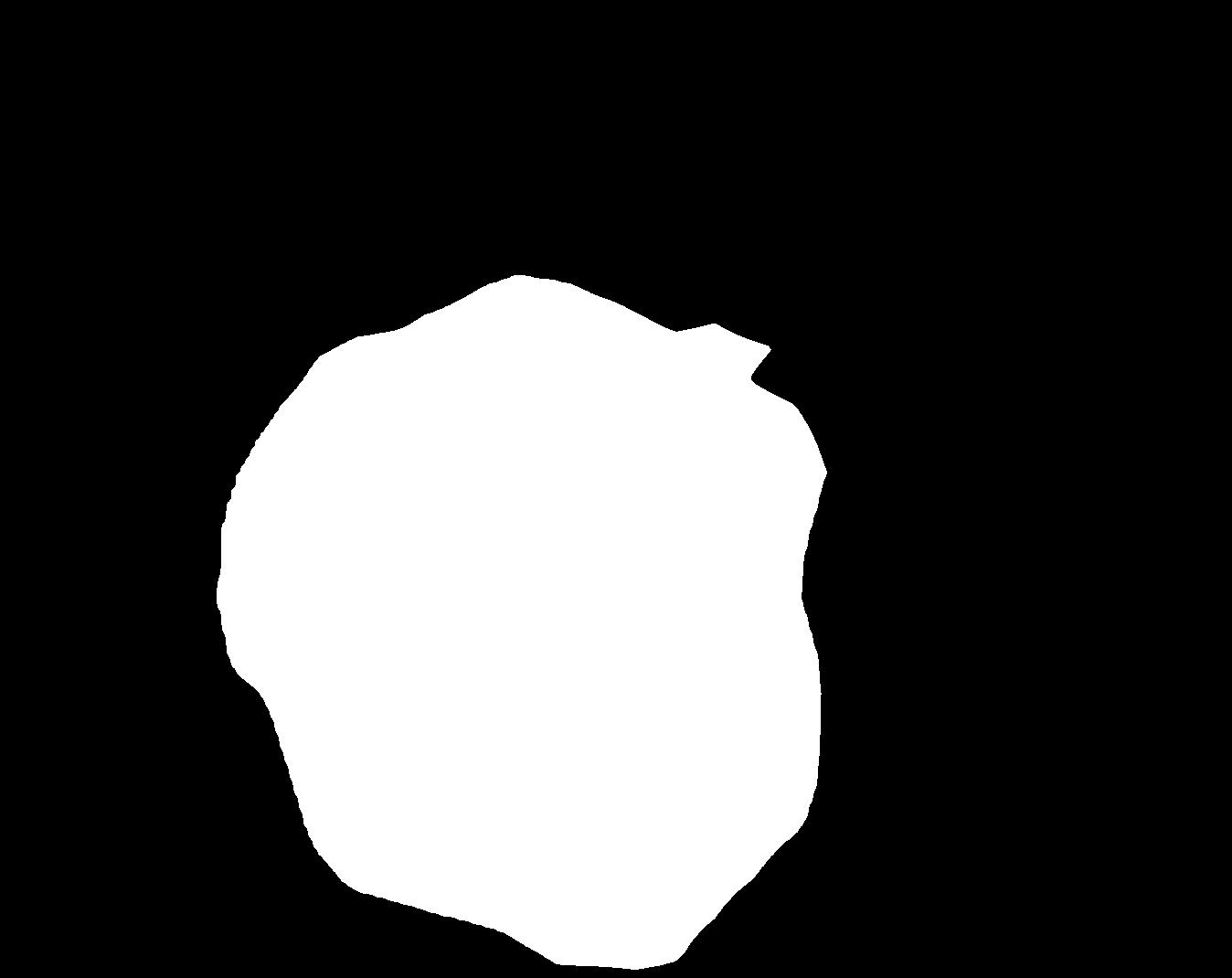} &
    \includegraphics[width=\imagesize\linewidth,height=\imagesize\linewidth]{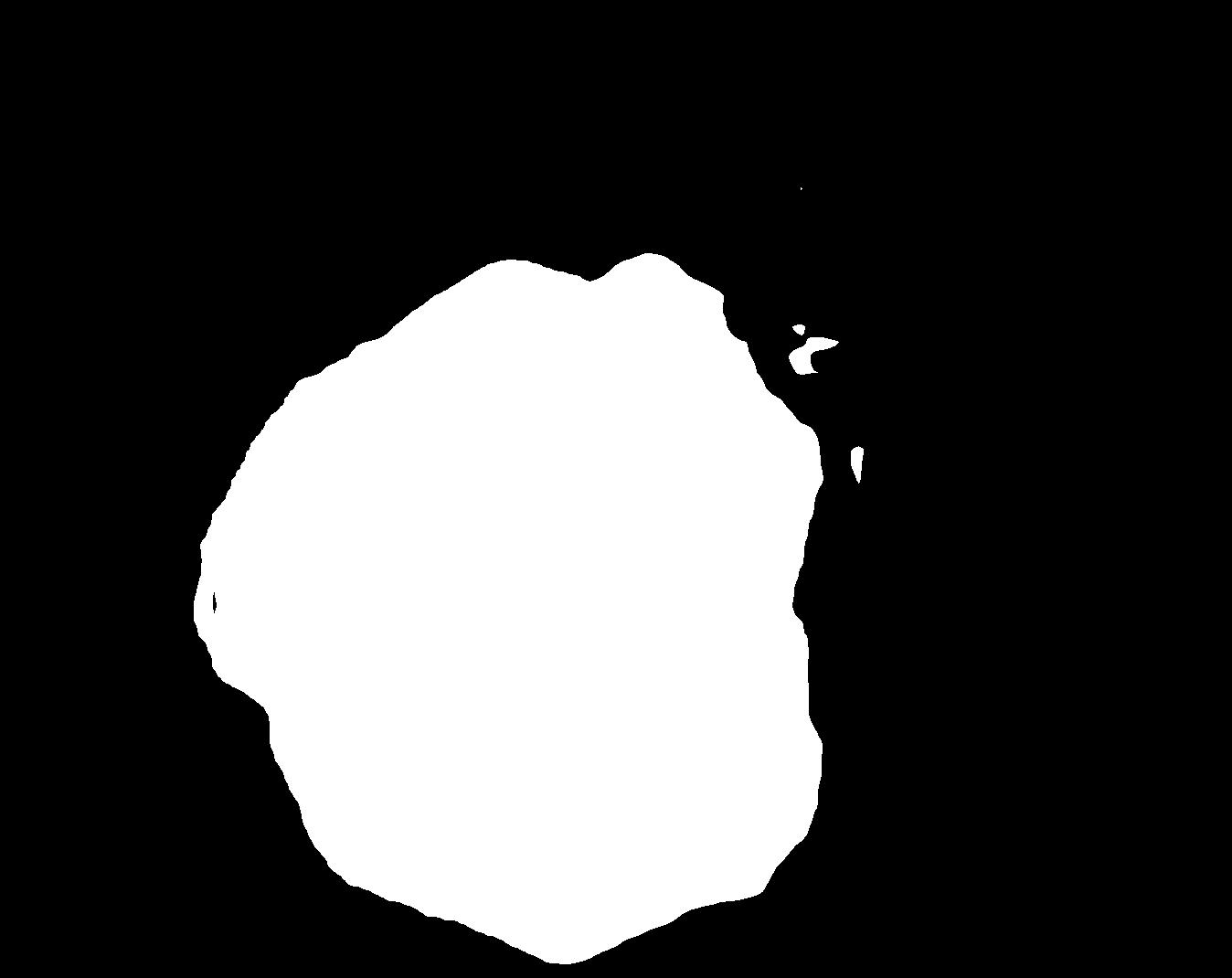} &
    \includegraphics[width=\imagesize\linewidth,height=\imagesize\linewidth]{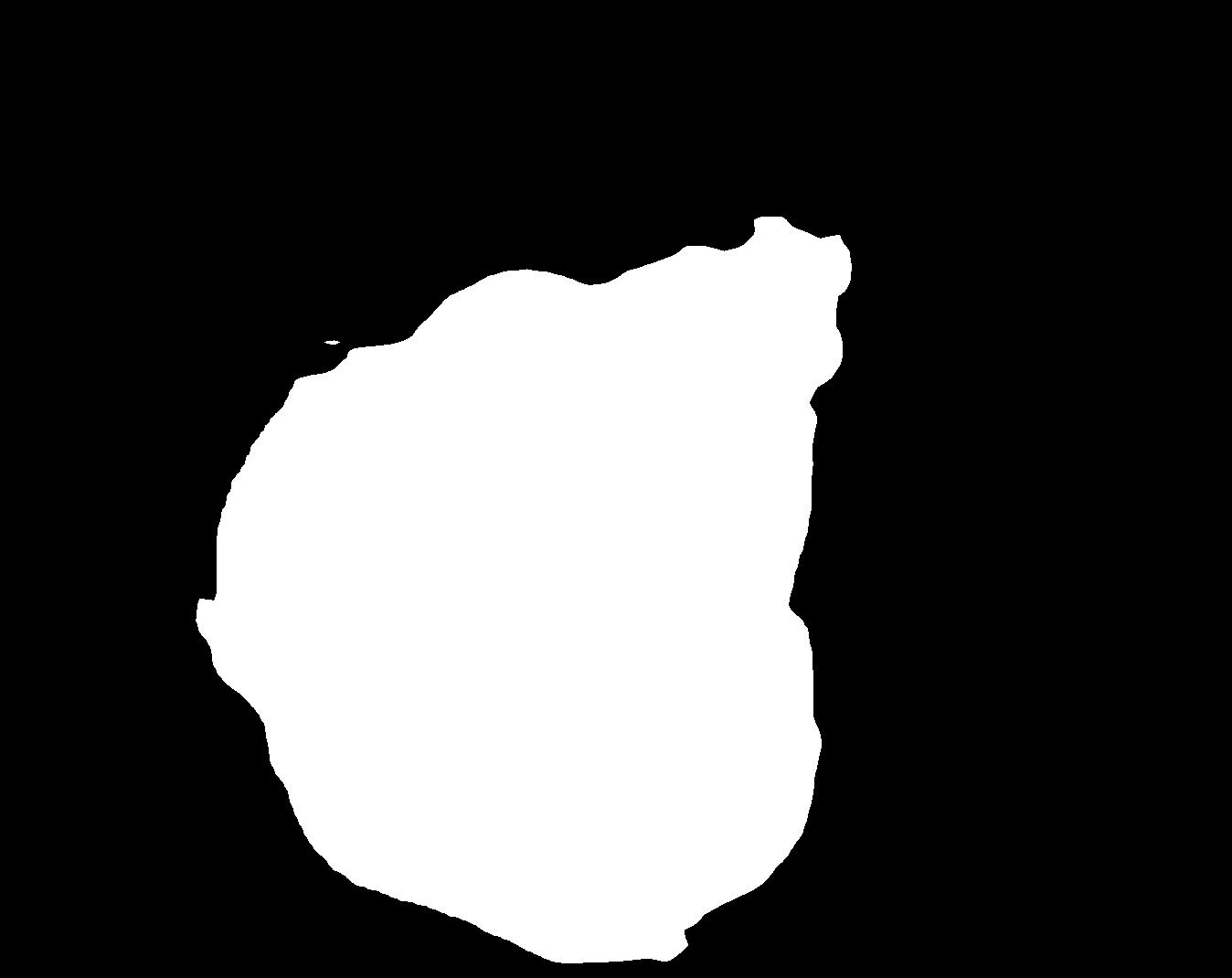} &
    \includegraphics[width=\imagesize\linewidth,height=\imagesize\linewidth]{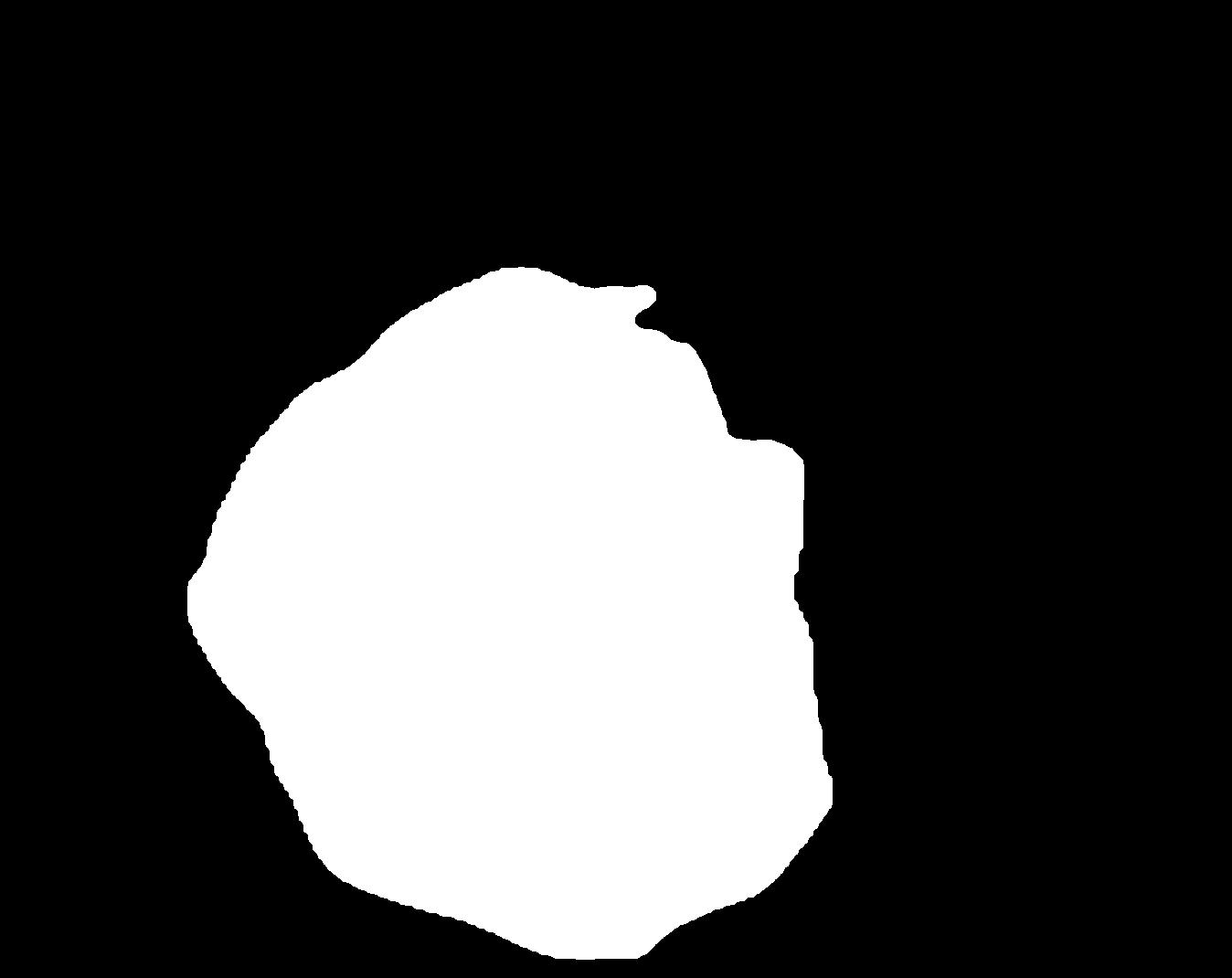} &
    \includegraphics[width=\imagesize\linewidth,height=\imagesize\linewidth]{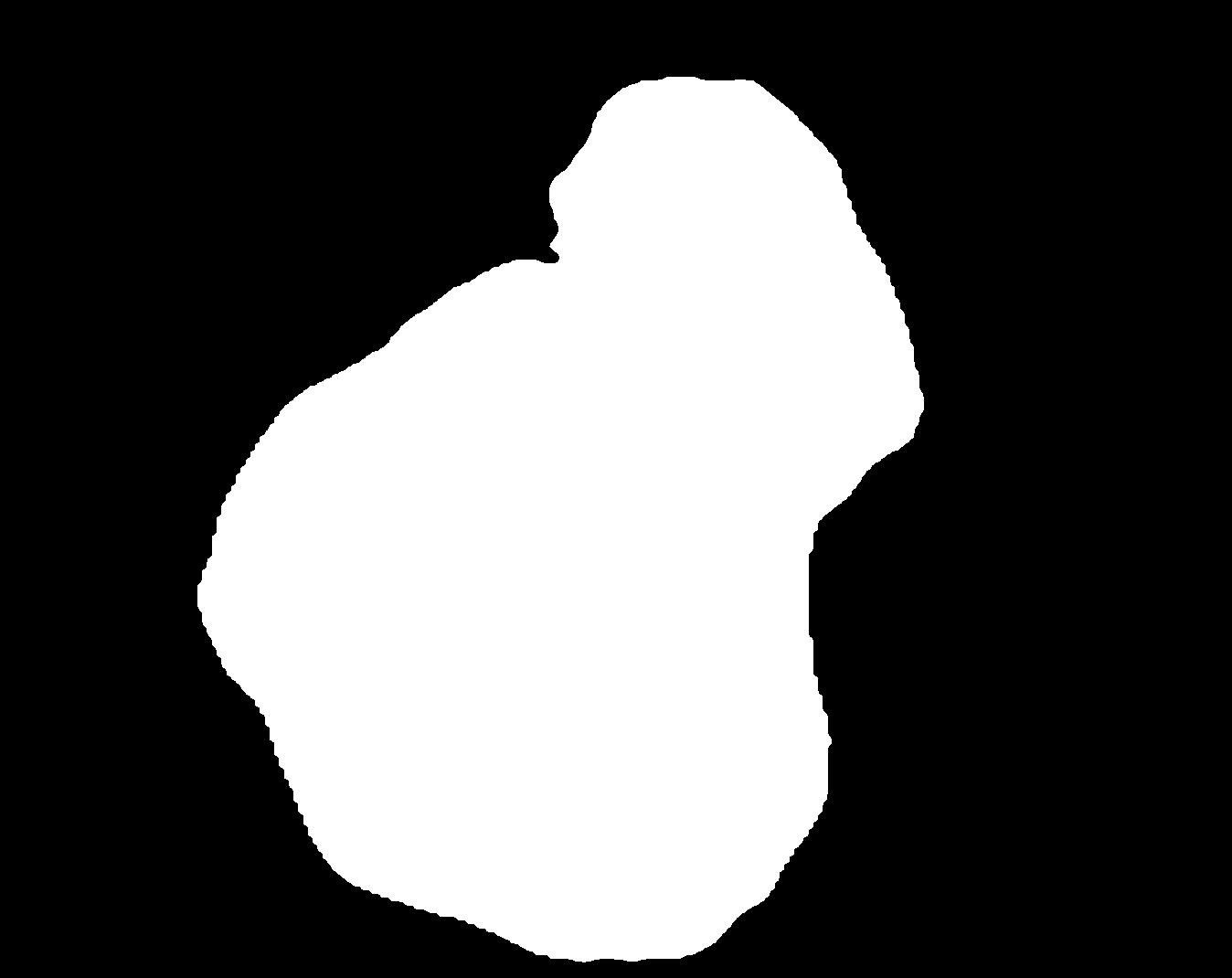} &
    \includegraphics[width=\imagesize\linewidth,height=\imagesize\linewidth]{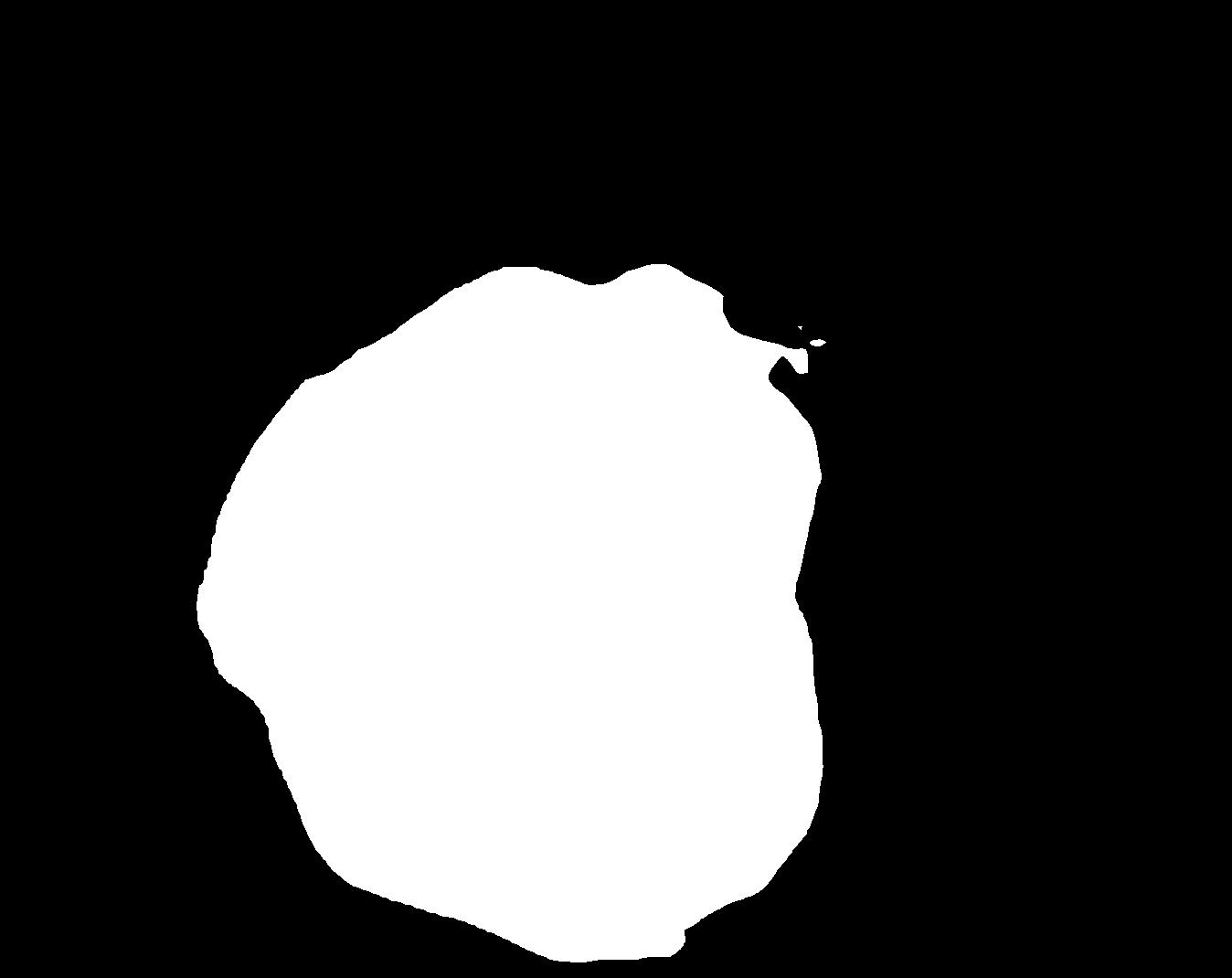} &
    \includegraphics[width=\imagesize\linewidth,height=\imagesize\linewidth]{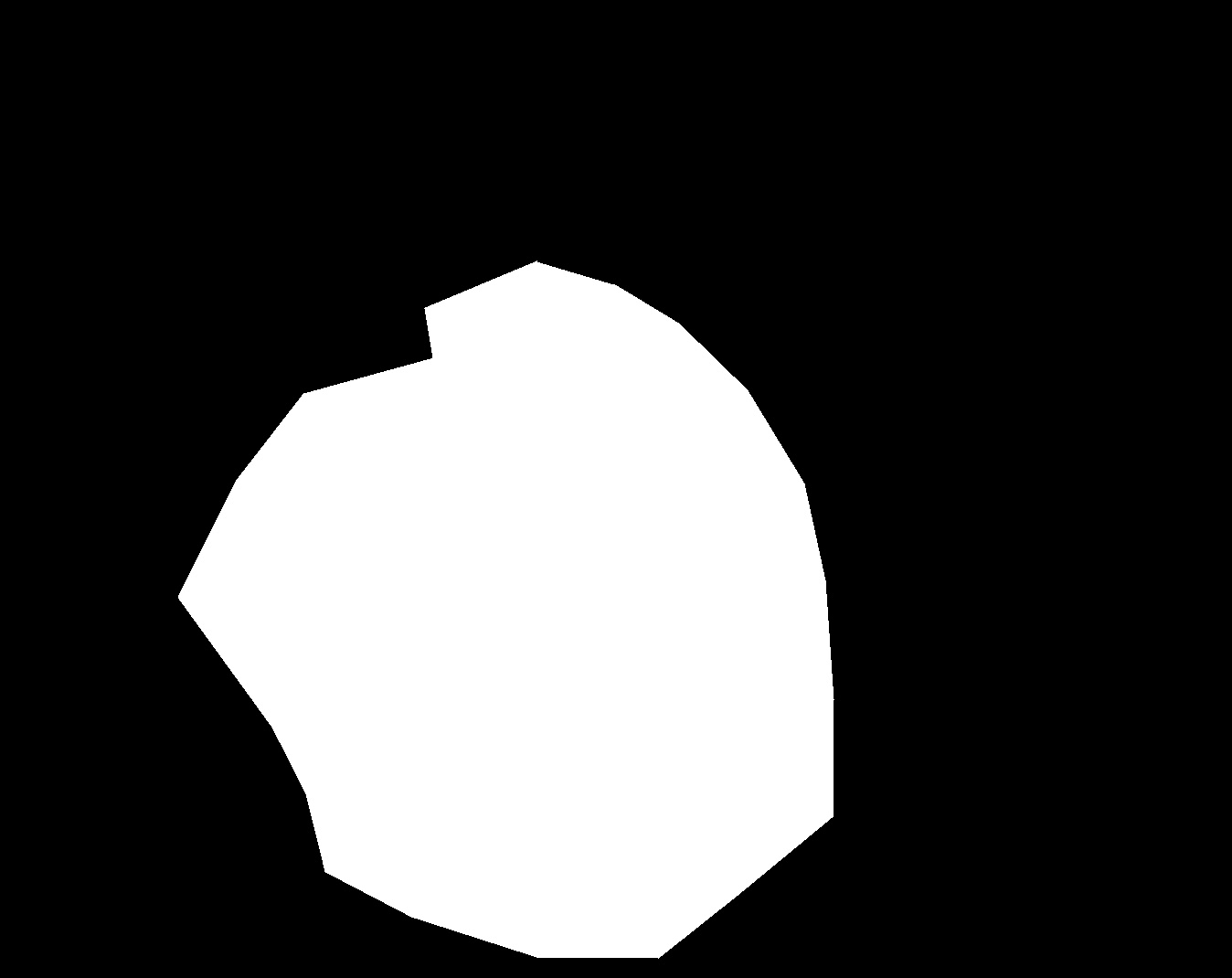} \\
    
    \includegraphics[width=\imagesize\linewidth,height=\imagesize\linewidth]{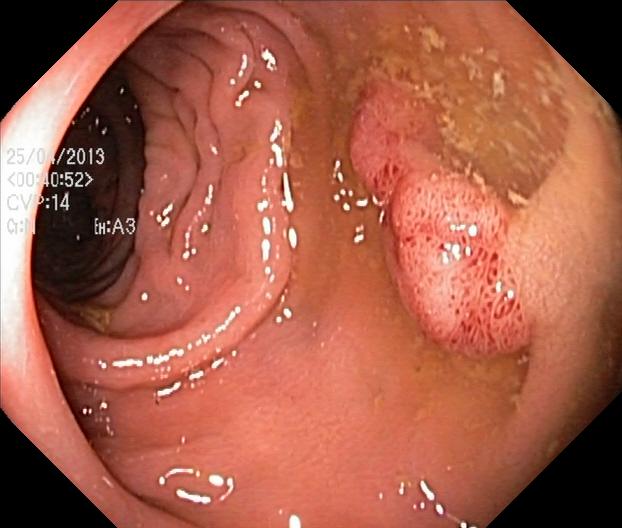} &
    \includegraphics[width=\imagesize\linewidth,height=\imagesize\linewidth]{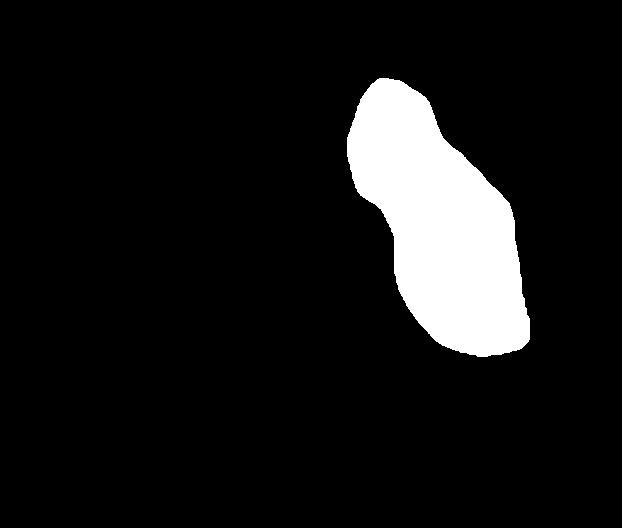} &
    \includegraphics[width=\imagesize\linewidth,height=\imagesize\linewidth]{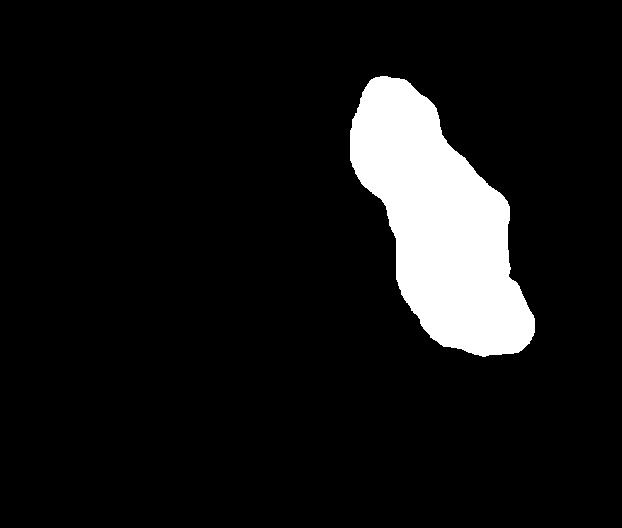} &
    \includegraphics[width=\imagesize\linewidth,height=\imagesize\linewidth]{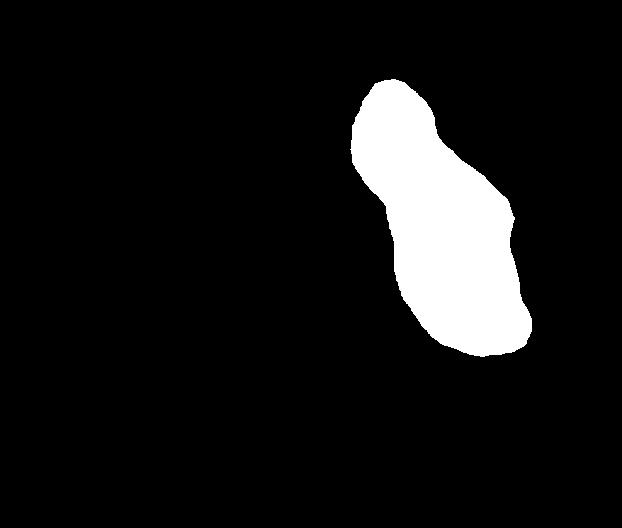} &
    \includegraphics[width=\imagesize\linewidth,height=\imagesize\linewidth]{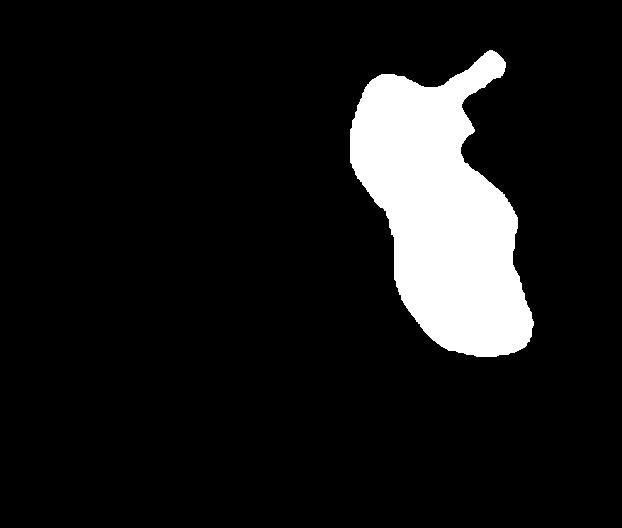} &
    \includegraphics[width=\imagesize\linewidth,height=\imagesize\linewidth]{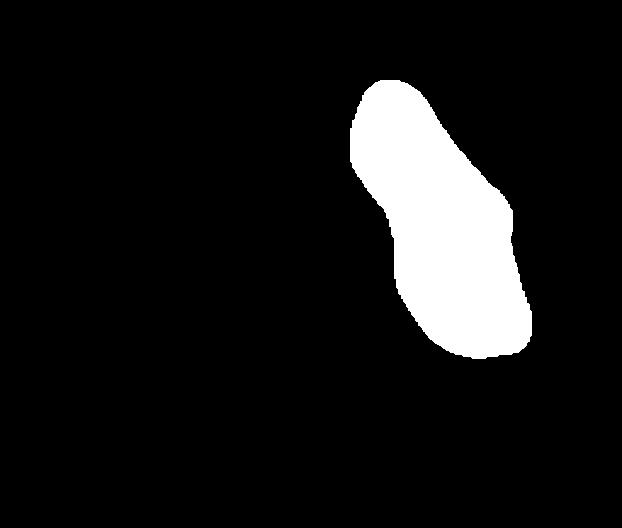} &
    \includegraphics[width=\imagesize\linewidth,height=\imagesize\linewidth]{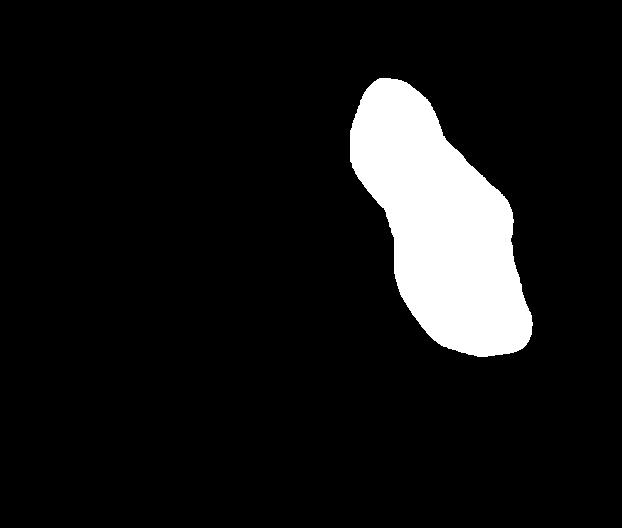} &
    \includegraphics[width=\imagesize\linewidth,height=\imagesize\linewidth]{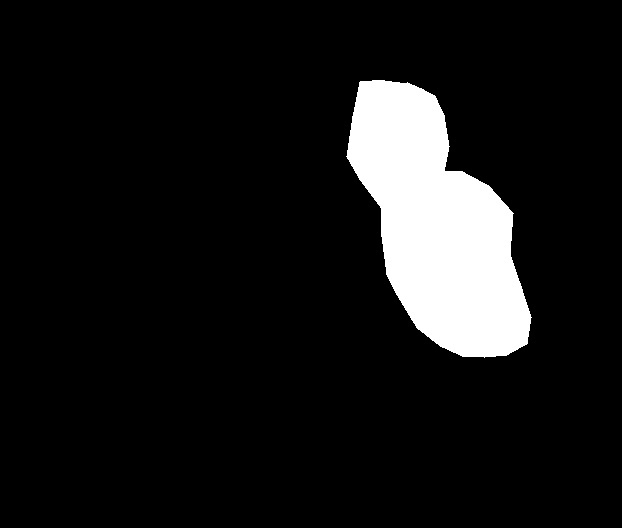} \\

    \includegraphics[width=\imagesize\linewidth,height=\imagesize\linewidth]{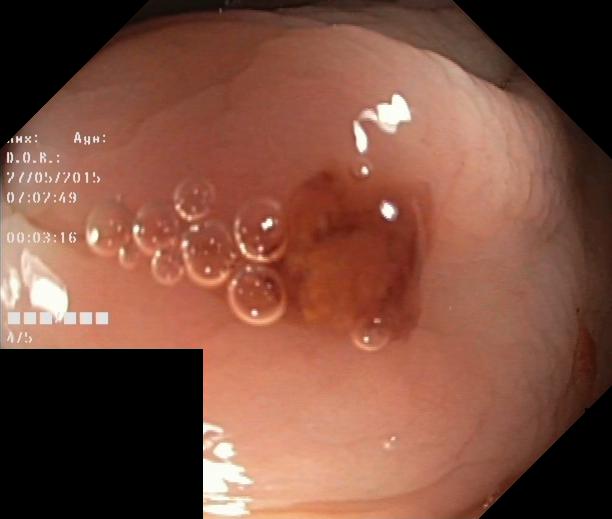} &
    \includegraphics[width=\imagesize\linewidth,height=\imagesize\linewidth]{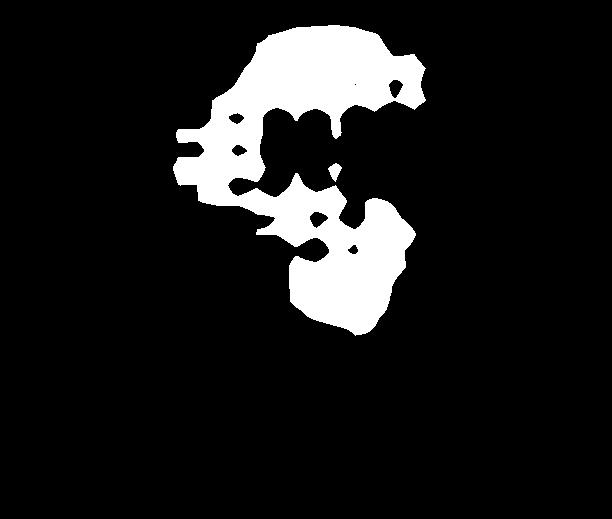} &
    \includegraphics[width=\imagesize\linewidth,height=\imagesize\linewidth]{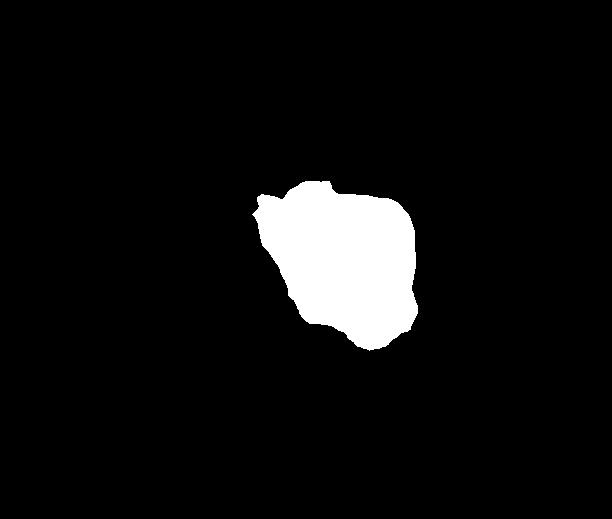} &
    \includegraphics[width=\imagesize\linewidth,height=\imagesize\linewidth]{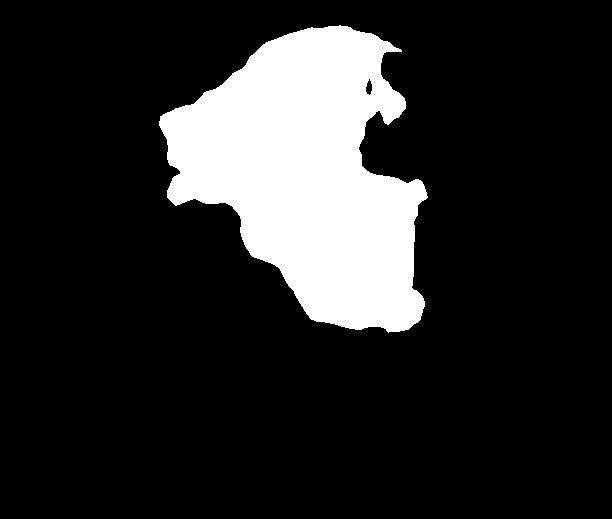} &
    \includegraphics[width=\imagesize\linewidth,height=\imagesize\linewidth]{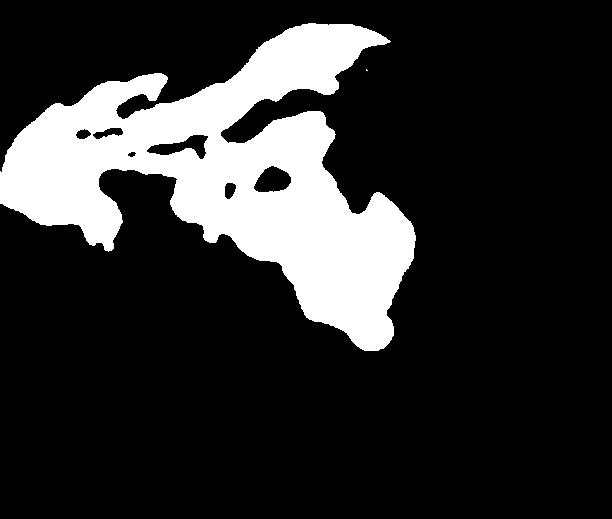} &
    \includegraphics[width=\imagesize\linewidth,height=\imagesize\linewidth]{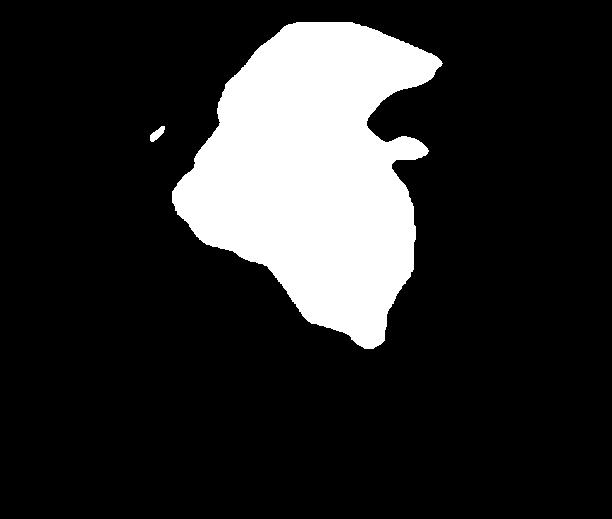} &
    \includegraphics[width=\imagesize\linewidth,height=\imagesize\linewidth]{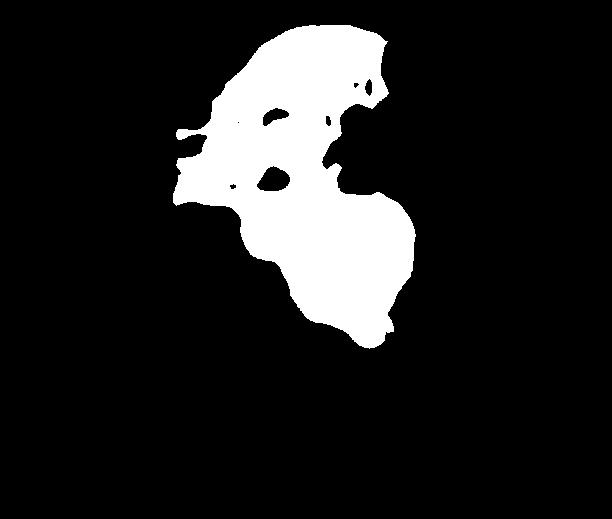} &
    \includegraphics[width=\imagesize\linewidth,height=\imagesize\linewidth]{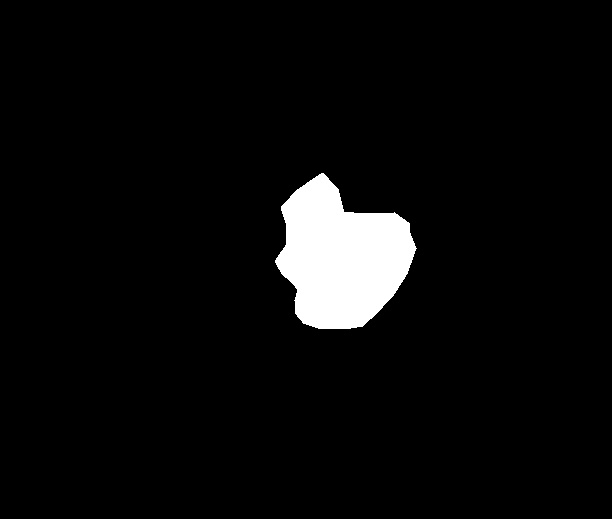} \\
    
    \includegraphics[width=\imagesize\linewidth,height=\imagesize\linewidth]{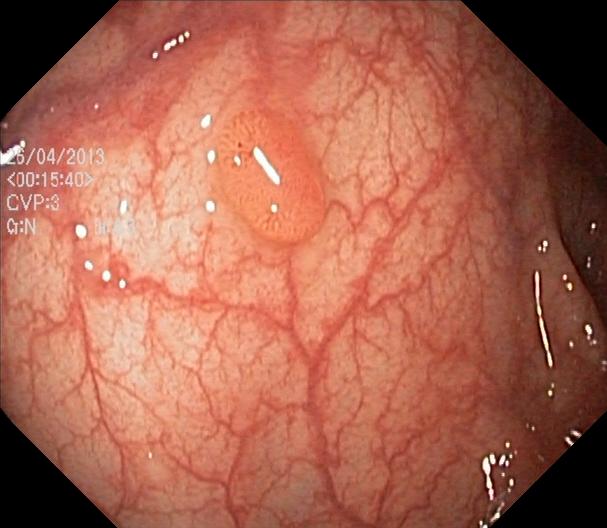} &
    \includegraphics[width=\imagesize\linewidth,height=\imagesize\linewidth]{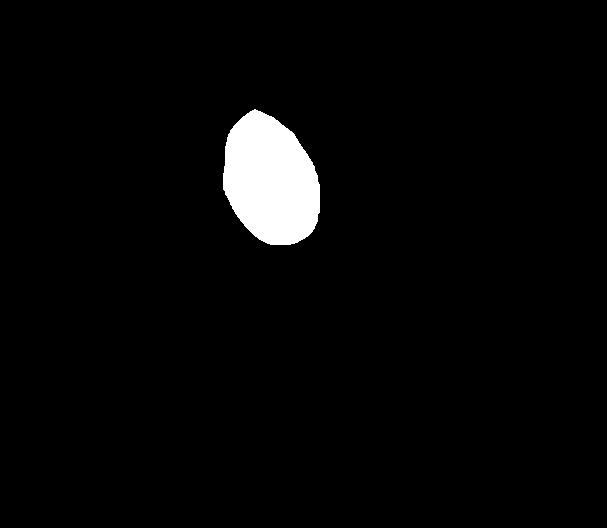} &
    \includegraphics[width=\imagesize\linewidth,height=\imagesize\linewidth]{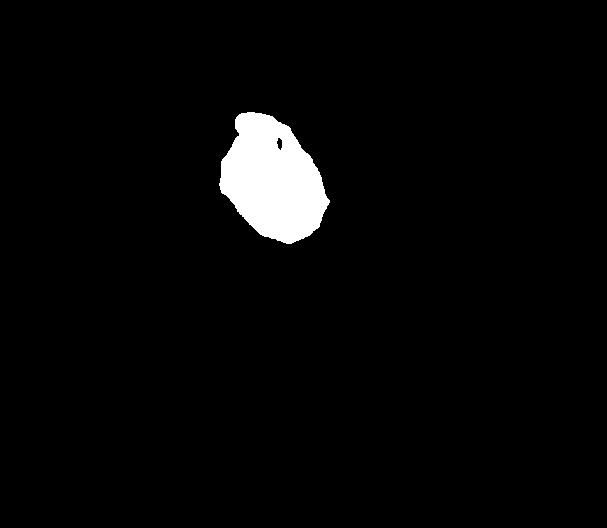} &
    \includegraphics[width=\imagesize\linewidth,height=\imagesize\linewidth]{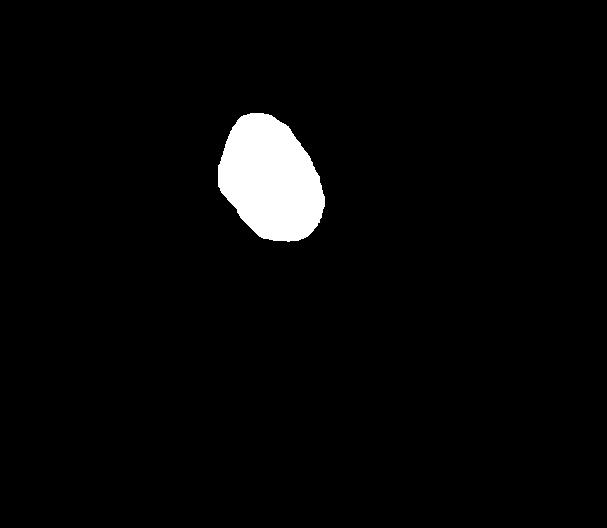} &
    \includegraphics[width=\imagesize\linewidth,height=\imagesize\linewidth]{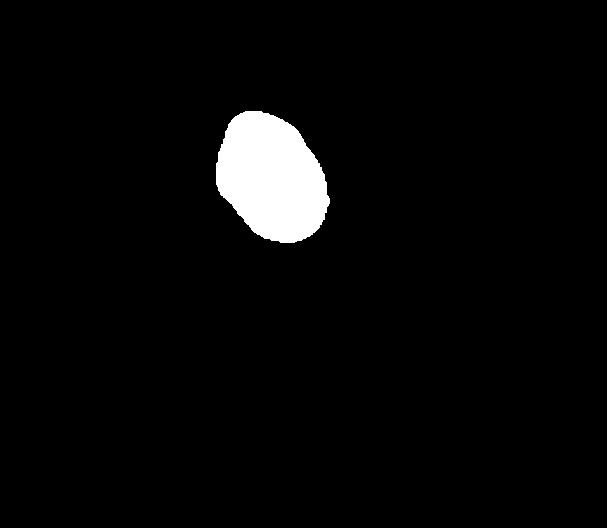} &
    \includegraphics[width=\imagesize\linewidth,height=\imagesize\linewidth]{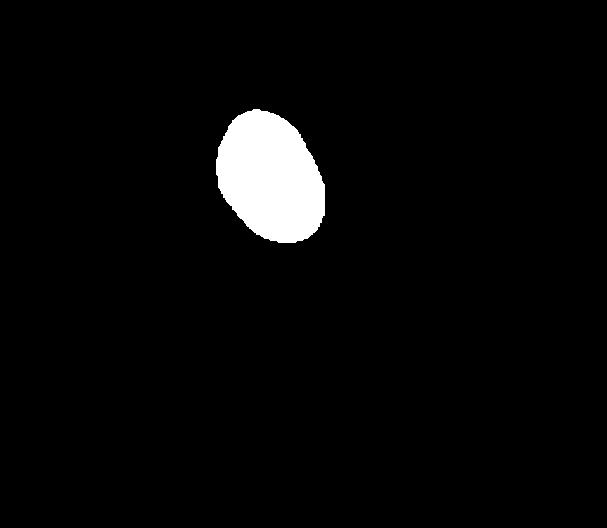} &
    \includegraphics[width=\imagesize\linewidth,height=\imagesize\linewidth]{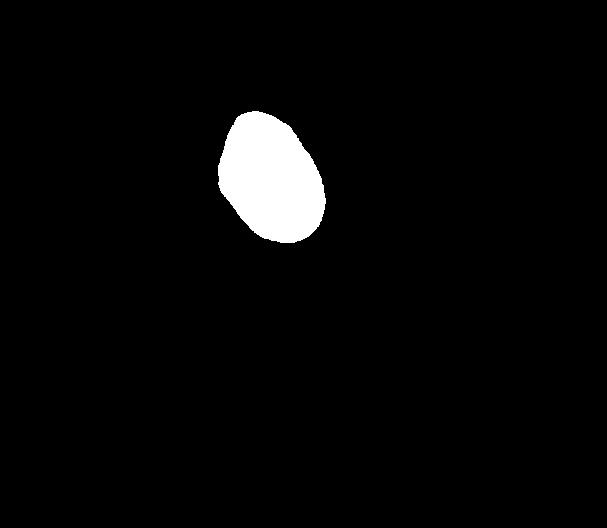} &
    \includegraphics[width=\imagesize\linewidth,height=\imagesize\linewidth]{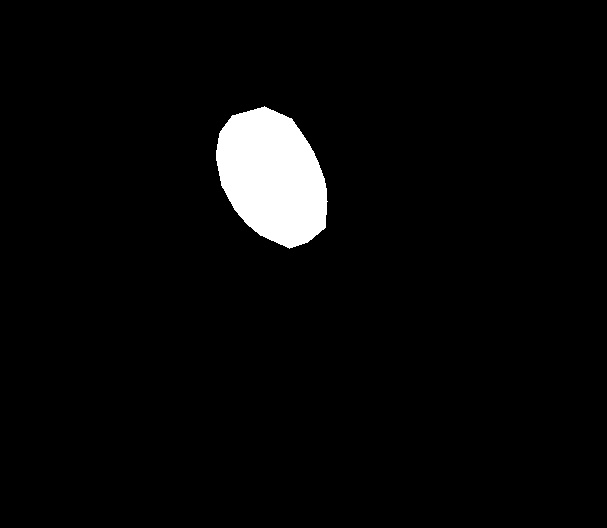} \\
    
    \includegraphics[width=\imagesize\linewidth,height=\imagesize\linewidth]{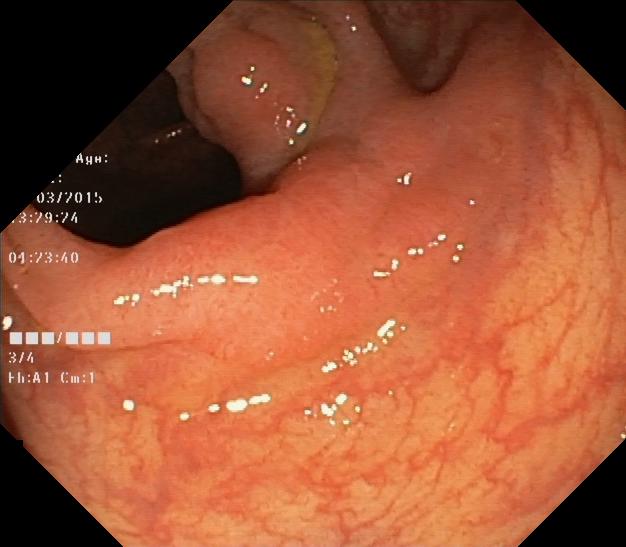} &
    \includegraphics[width=\imagesize\linewidth,height=\imagesize\linewidth]{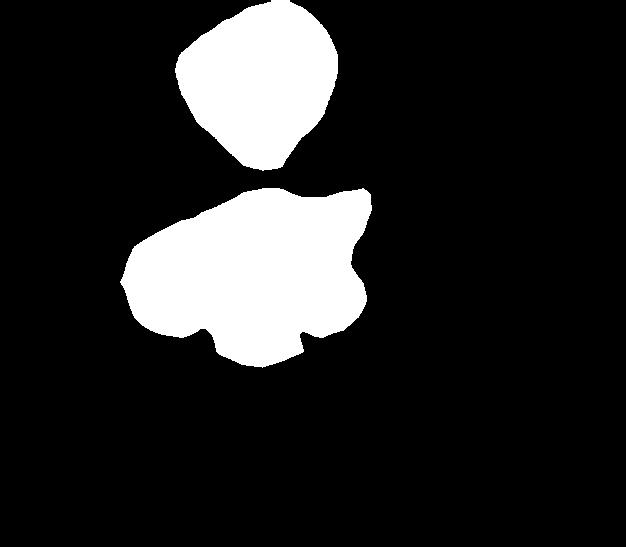} &
    \includegraphics[width=\imagesize\linewidth,height=\imagesize\linewidth]{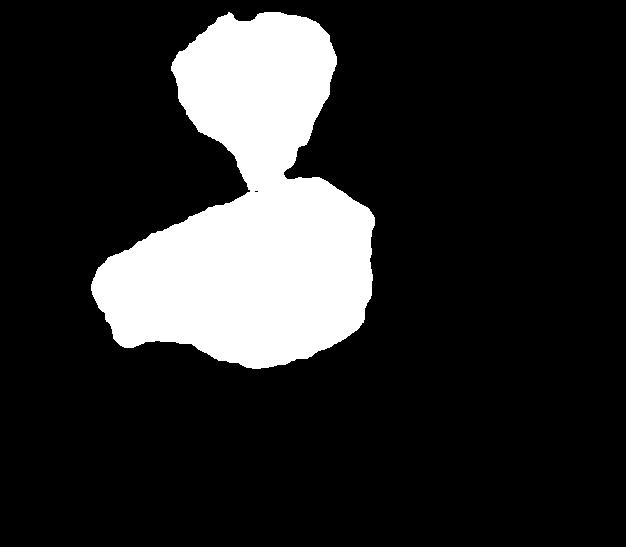} &
    \includegraphics[width=\imagesize\linewidth,height=\imagesize\linewidth]{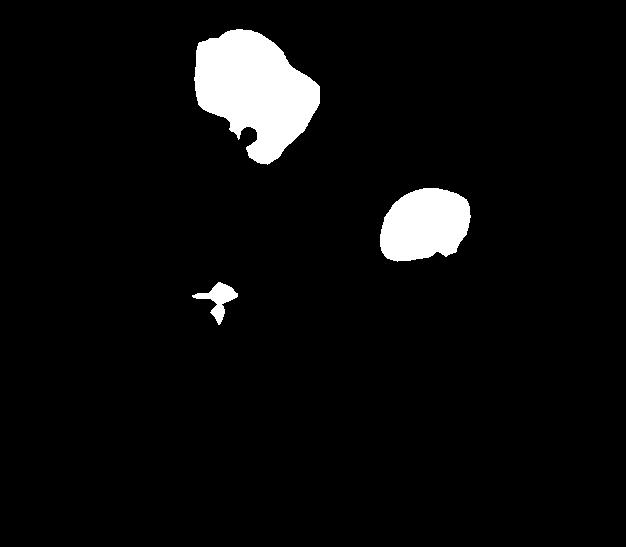} &
    \includegraphics[width=\imagesize\linewidth,height=\imagesize\linewidth]{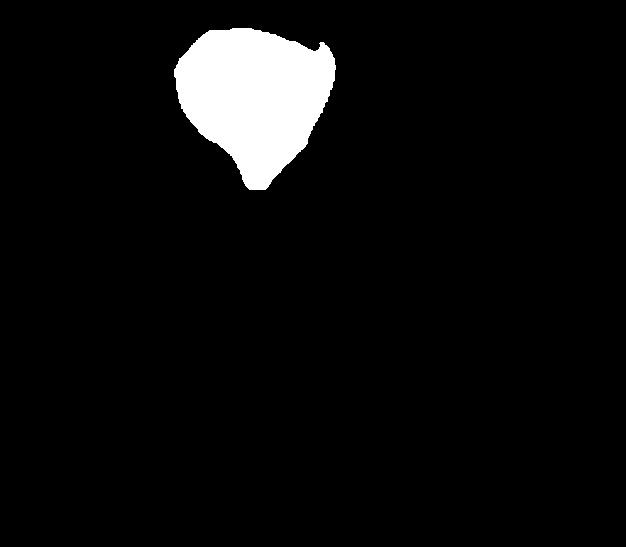} &
    \includegraphics[width=\imagesize\linewidth,height=\imagesize\linewidth]{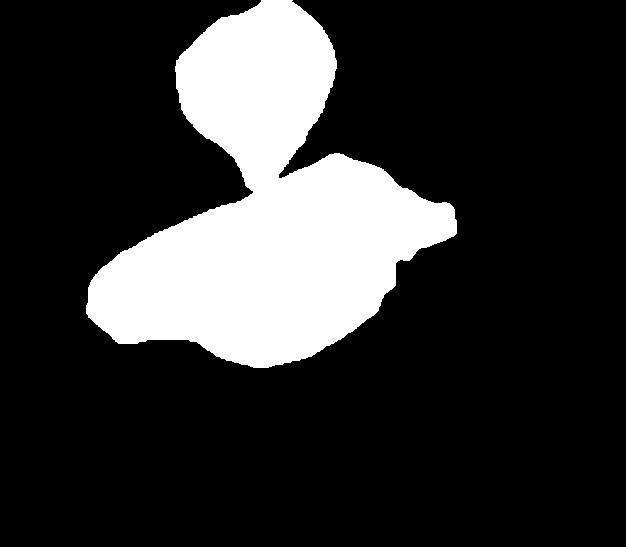} &
    \includegraphics[width=\imagesize\linewidth,height=\imagesize\linewidth]{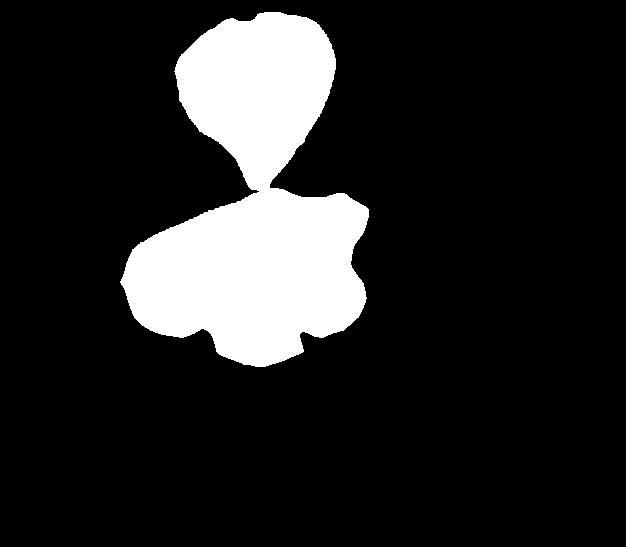} &
    \includegraphics[width=\imagesize\linewidth,height=\imagesize\linewidth]{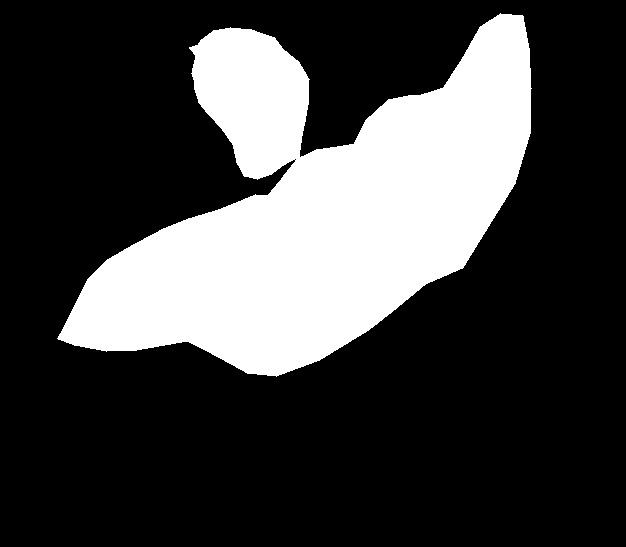} \\

\end{tabular}
    \caption{Some predicted mask examples taken from the divergent network and its five intermediate models. The images are taken from HyperKvasir.}\label{fig:predicted_masks}
\end{figure}